\documentclass[aps,prd,showpacs,groupedaddress,superscriptaddress,%
floatfix,nofootinbib,twocolumn]{revtex4}%

\usepackage{amsmath}
\usepackage{graphicx}
\usepackage{subfigure}
\usepackage{epstopdf}
\usepackage{epsfig}
\usepackage{amssymb}
\usepackage{bm}
\usepackage{bbm}
\newcommand{\beq}{\begin{equation}}
\newcommand{\eeq}{\end{equation}}

\usepackage{color}

\begin{document}

\title{ 
Top $\bm{A_{FB}}$ at the Tevatron vs. charge asymmetry at the LHC
\\ in chiral U(1) flavor models with flavored Higgs doublets
}

\author{P. Ko}
\affiliation{School of Physics, KIAS, Seoul 130-722, Korea}

\author{Yuji Omura}
\affiliation{School of Physics, KIAS, Seoul 130-722, Korea}

\author{Chaehyun Yu}
\affiliation{School of Physics, KIAS, Seoul 130-722, Korea}


\begin{abstract}
\noindent
We consider the top forward-backward (FB) asymmetry at the Tevatron and
top charge asymmetry at the LHC within  chiral U(1)$^\prime$ models with
flavor-dependent U(1)$^\prime$ charges and flavored Higgs fields,
which were introduced in the Ref.~\cite{u1models2}.  In this model, one has 
to include the flavor changing couplings of the Higgs bosons as well as the
Z$^\prime$ to the up-type quarks. 
The models could enhance not only the top forward-backward asymmetry
at Tevatron, but also the top charge asymmetry at LHC, without too large
same-sign top-quark pair production rates. Also the $m_{t\bar{t}}$ distribution 
at high $m_{t\bar{t}}$ show less deviations from the SM predictions.
We identify parameter spaces for the U(1)$^{\prime}$ gauge boson
and (pseudo)scalar Higgs bosons where all the experimental data could be
accommodated, including the case with about $125$ GeV Higgs boson,
as suggested recently by ATLAS and CMS.
\end{abstract}


\maketitle

\section{Introduction \label{sec:intro}}

The top quark has been one of the most promising channels to search 
for new physics beyond the Standard Model (SM).
Among the observables involving the top quark, 
the top forward-backward asymmetry ($A_\textrm{FB}^t$) at the Tevatron
has been paid much attention during the past few years, because it is the only 
quantity which has some deviations from the SM prediction in top quark sector. 
The CDF Collaboration announced that  $A_\textrm{FB}^t$ 
in the $t\bar{t}$ rest frame 
is $( 0.158\pm 0.074 )$ in the lepton+jets channel~\cite{CDFlepjet}, and
$( 0.420\pm 0.158 )$ in the dilepton channel~\cite{CDFdilep}, respectively.  
The combined result of them is $( 0.201\pm 0.067 )$, which  is consistent
with $A_\textrm{FB}^t=0.196\pm 0.060^{+0.018}_{-0.026}$ announced 
by the D0 Collaboration in the lepton+jets channel~\cite{D0lepjet}.
Very recently, the CDF Collaboration updated the results for $A_\textrm{FB}^t$ 
in the lepton+jets channel with data of a luminosity of $8.7$ fb$^{-1}$: $A_\textrm{FB}^t=0.162\pm0.047$~\cite{cdfnew}.
This value is consistent with the previous measurements at CDF and D0.
The SM predictions for $A_\textrm{FB}^t$ are $0.072^{+0.011}_{-0.007}$ 
at the next-to-leading order (NLO) + next-to-next-to-leading logarithm
accuracies~\cite{smnlo,smnlo2} and $0.087\pm 0.010$ with NLO corrections
for the electroweak interactions~\cite{smewnlo,smewnlo2}, respectively.  
Therefore there is still about 2$\sigma$ deviation between the SM prediction
and experiments in the integrated $A_{\textrm{FB}}^t$ at the Tevatron.

A number of new models have been proposed to account for the discrepancy 
in $A_\textrm{FB}^t$~\cite{Choudhury:2007ux,Jung:2009jz,Cheung:2009ch,%
Shu:2009xf,Arhrib:2009hu,Dorsner:2009mq,Jung:2009pi,Barger:2010mw,Cao:2010zb,%
Xiao:2010hm,Jung:2010yn,Choudhury:2010cd,Cheung:2011qa,Gresham:2011dg,%
Bhattacherjee:2011nr,Barger:2011ih,Grinstein:2011yv,%
Patel:2011eh,Isidori:2011dp,%
Zerwekh:2011wf,Barreto:2011au,Foot:2011xu,Ligeti:2011vt,AguilarSaavedra:2011vw,%
Gresham:2011pa,Shu:2011au,AguilarSaavedra:2011zy,Nelson:2011us,Jung:2011ua,%
Zhu:2011ww,Jung:2011ue,Babu:2011yw,Krohn:2011tw,Hektor:2011ms,Cui:2011xy,%
Gabrielli:2011jf,Duraisamy:2011pt,AguilarSaavedra:2011ug,Tavares:2011zg,%
Vecchi:2011ab,Shao:2011wa,Blum:2011fa,Gresham:2011fx,Frank:2011rb,%
Davoudiasl:2011tv,Jung:2011id,Liu:2011dh,Kolodziej:2011ir,Ng:2011jv,Yan:2011tf,%
Jung:2011ym,Wang:2011mra,Biswal:2012mr,Ko:2012sm,Gresham:2012wc,%
Grinstein:2012pn,Han:2012qu,Duffty:2012zz,%
u1models,u1models2,u1models3}.
In order that those new models can accommodate the present data 
in the Drell-Yan, dijet production, flavor-changing-neutral-current (FCNC)
experiments and so on, it is usually assumed that the new model has 
a large coupling only to the top quark. 
In general, it would be a challenging task to construct 
a realistic and consistent model with such a hierarchy in couplings.
New models should also be anomaly-free and have proper Yukawa interactions.
Otherwise there could be some hidden fields that might affect 
the physical observables we are interested in. 

One of the most interesting models to account for $A_\textrm{FB}^t$
is a $Z^\prime$ model with an additional chiral U(1)$^\prime$ symmetry,
where only the right-handed (RH) up-type quarks in the SM are 
charged under the U(1)$^\prime$ symmetry~\cite{u1models,u1models2,u1models3}. 
Such a chiral U(1)$^\prime$ symmetry is inevitably 
accompanied with the modification of the Higgs sector in the SM, 
for instance, the addition of the multi-Higgs doublets 
charged under the U(1)$^\prime$ symmetry. There are two reasons why
we have to extend the Higgs sector. 

First of all, all the SM fermions charged under chiral U(1)$^\prime$ 
(such as top quark in Refs.~\cite{u1models,u1models2}) 
would be massless and so the model becomes 
unphysical without extra Higgs doublets charged under U(1)$^\prime$. 
There is no limit where one can integrate out these U(1)$^\prime$-charged Higgs 
doublets assuming they are heavy. 
Secondly, the mass of the $Z^\prime$ boson could be generated through 
the U(1)$^\prime$ breaking. One can achieve this either by the  
U(1)$^\prime$-charged Higgs doublets or U(1)$^\prime$-charged Higgs singlet.
Without these Higgs fields that generate the $Z^\prime$ mass, the theory 
violates unitarity at high energy, and it loses its predictability.
This situation is somewhat similar to the unitarity problem of the 
$W_L W_L \to W_L W_L$ or $f \bar{f} \rightarrow W_L W_L$ scatterings 
in the intermediate vector boson model. 
The unitarity of the model would be restored with the U(1)$^\prime$-charged
Higgs fields (doublets and singlet). 
That is, additional Higgs fields are mandatory for generation
of masses of the SM fermions charged under chiral U(1)$^\prime$ and the 
U(1)$^\prime$ gauge boson ($Z^\prime$) itself. 
The requirement for the extra Higgs fields  is also true 
in the $W^\prime$~\cite{Cheung:2009ch,Babu:2011sd},
axigluon~\cite{Tavares:2011zg}, flavor SU(3)$_{u_R}$~\cite{Grinstein:2011yv},
and any other models if the  SM fermions are chiral under new gauge interactions.

On the other hand, the gauge anomaly would exist in general 
if one introduces the flavor-dependent chiral U(1)$^\prime$ symmetry. 
But  it could be canceled by introducing extra fermions~\cite{u1models,u1models2}.
We emphasize that in order to construct a realistic model for the chiral 
U(1)$^\prime$ symmetry, one must take into account carefully  
the U(1)$^\prime$-charged Higgs fields and extra fermion fields in addition to 
the extra U(1)$^\prime$ gauge field. 
It is important to realize that there is no proper limit where 
only light $Z^{'}$ survives with chiral couplings to the right-handed 
up-type quarks. 
Literally speaking, the model by Jung, Murayama, Pierce 
and Wells~\cite{Jung:2009jz} is not well defined as a gauge theory and 
not realistic, since the up-type quarks are all massless including top quarks. 
It is mandatory to extend the model by introducing  extra Higgs doublets  
with nonzero U(1)$^\prime$ charges, because the up-type quarks 
(including top quark) would be all massless  without the extra Higgs doublets.

The Large Hadron Collider (LHC) is also called a ``Top Factory'' because 
a huge number of top-quark pair (larger than about $10^6$) is expected to be 
produced at the LHC even before the technical shutdown at the end of this year. 
Thus  new models proposed for resolutions of $A_\textrm{FB}^t$ 
at the Tevatron including the flavor-dependent chiral U(1)$^\prime$ model 
with flavored Higgs doublets~\cite{u1models,u1models2} could be tested 
at the LHC.  For instance, the original $Z^\prime$ model 
is strongly constrained by the same-sign top-quark pair production 
at the LHC~\cite{Chatrchyan:2011dk,Aad:2012bb,Cao:2011ew,Berger:2011ua}.

Another interesting observable at the LHC is  the top charge asymmetry $A_C^y$ 
defined by the difference of numbers of events  with the positive and negative 
$\Delta |y|$ divided by their sum: 
\begin{equation}
A_C^y = \frac{N(\Delta |y|>0) - N(\Delta |y|<0)}
{N(\Delta |y|>0) + N(\Delta |y|<0)},
\label{ac}%
\end{equation}
where $\Delta |y| = |y_t|-|y_{\bar{t}}|$ for the rapidities $y_{t}$
and $y_{\bar{t}}$ of the top and anti-top quarks. Because the LHC is 
a symmetric collider under charge conjugation, $A_\textrm{FB}^t$ cannot
be defined at the LHC, unlike the Tevatron. 
However, at the LHC the top quark produced in the $q\bar{q}\to t\bar{t}$ 
process is statistically more boosted to the beam direction compared to 
the anti-top quark because the top quark follows the direction of the 
incident quark which has a larger longitudinal momentum. 
This difference could generate the charge asymmetry in Eq.~(\ref{ac}).
The theoretical estimate for $A_C^y$ is about $0.01$ at NLO~\cite{smnlo},
which is consistent with the empirical data
$A_C^y = -0.018\pm 0.028\pm 0.023$ at ATLAS~\cite{atlasacy}
and $A_C^y = 0.004\pm 0.010 \pm 0.012$ at CMS~\cite{cmsacy}
within uncertainties.

In the previous works~\cite{u1models,u1models2,u1models3}, we considered 
only  $A_{\textrm{FB}}^t$ at the Tevatron and the same-sign top-quark pair 
production  rate at the LHC in order to find parameter regions of the models 
which are consistent with the data available at that time. 
In this paper, we consider top charge asymmetry at the LHC within 
the same chiral U(1)$^\prime$ model, 
taking into account more stringent recent constraints on the same-sign 
top-quark pair production at ATLAS~\cite{Aad:2012bb}, 
and investigate if the model in Refs.~\cite{u1models,u1models2,u1models3} 
survives or not, and how we can test further these models at the LHC.

This paper is organized as follows. In Sec.~\ref{sec:model}, we briefly
review the chiral U(1)$^\prime$ models with flavored Higgs doublets
which were first proposed in Refs.~\cite{u1models,u1models2}, 
with focus on the Lagrangian relevant to $A_\textrm{FB}^t$ at the Tevatron and 
$A_C^y$ at the LHC.  In Sec.~\ref{sec:pheno},  we discuss the phenomenology of 
our model at the Tevatron and LHC.   Finally we conclude in Sec.~\ref{sec:con}.

\section{Chiral U(1)$^{'}$ model with flavored Higgs doublets \label{sec:model}}

In this section we review the flavor-dependent chiral U(1)$^\prime$ model 
with flavored Higgs doublets that were proposed in Refs.~\cite{u1models,u1models2}. 
Our model is an extension of a simple phenomenological $Z^\prime$ model  with a 
large flavor changing neutral couplings in the $u_R - t_R$ sector~\cite{Jung:2009jz}. 
This $Z^\prime$ boson must be associated with some gauge symmetry 
if we work in weakly interacting theories. 
As a simple example we considered an extra U(1)$^\prime$ symmetry
~\cite{u1models,u1models2}.  The $Z^\prime$ boson better be leptophobic to avoid
the stringent constraints from the LEP II and Drell-Yang experiments.
Furthermore, it would be very difficult to assign flavor-dependent
U(1)$^\prime$ charges to the down-type quarks and left-handed up-type quarks
because it gives rise to dangerous FCNCs.
Therefore we assigned flavor-dependent  U(1)$^\prime$ charges $u_i$ $(i=u,c,t)$
only to the right-handed up-type quarks while the left-handed quarks and
right-handed down-type quarks have universal charges $Q_L$ and $d_R$ 
under U(1)$^\prime$. For simplicity, we assign $Q_L=d_R=0$.

Then, the Lagrangian between $Z^\prime$ and the SM quarks 
in the interaction eigenstates is given by
\begin{equation}
\mathcal{L}_{Z^\prime q\bar{q}} =
g^\prime \sum_i u_i  Z^\prime_\mu \overline{U_R^i} \gamma^\mu U_R^i,
\end{equation}
where $U_R^i$ is a right-handed up-type quark field in the interaction eigenstates 
and $g^\prime$ is the couping of the U(1)$^\prime$. 

After the electroweak symmetry breaking, we can rotate the quark fields 
into the mass eigenstates by bi-unitary transformation.   
The interaction Lagrangian for the $Z^\prime$ boson in the mass eigenstate
is given by
\begin{widetext}
\begin{equation}
\mathcal{L}_{Z^\prime q\bar{q}} =
g^\prime Z^\prime_\mu 
\left[ 
(g_R^u)_{ut} \overline{u_R} \gamma^\mu t_R
+(g_R^u)_{ut} \overline{t_R} \gamma^\mu u_R
+(g_R^u)_{uu} \overline{u_R} \gamma^\mu u_R
+(g_R^u)_{tt} \overline{t_R} \gamma^\mu t_R
\right].
\end{equation}
\end{widetext}
The $3\times 3$ mixing matrix $(g_R^u)_{ij} = (R_u)_{ik} u_k (R_u)^\dagger_{kj}$ 
is the product of the U(1)$^\prime$ charge matrix ${\rm diag} ( u_{k=1,2,3} )$ and
a unitary matrix $R_u$, where the matrix  $R_u$ relates the RH up-type quarks 
in the interaction eigenstates and in the mass eigenstates.   
The matrix $R_u$ participates in diagonalizing the up-type quark 
mass matrix. In principle, the mixing matrix $g_R$ could be complex, providing  
an additional source of CP violation in the right-handed up-type quark sector. 
In this work, we assume it is real for simplicity. 
We note that the components of the mixing angles related 
to the charm quark have to be small in order to respect 
constraints from the $D^0$-$\overline{D^0}$ mixing.

If one assigns the U(1)$^\prime$ charge $(u_i)=(0,0,1)$
to the right-handed up-type quarks, one can find the relation 
$(g_R^u)_{ut}^2 = (g_R^u)_{uu} (g_R^u)_{tt}$ \footnote{We note 
that the relation is not valid for the other charge assignments.}. 
This relation indicates that if the $t$-channel diagram mediated by $Z^\prime$ 
contributes to the $u\bar{u}\to t\bar{t}$ process, the $s$-channel diagram 
mediated by $Z^\prime$ should be taken into account, too.

As we discussed in the previous section, it is mandatory to include
additional flavored Higgs doublets charged under U(1)$^\prime$ in order 
to write down proper Yukawa interactions for the SM quarks charged
under U(1)$^\prime$ at the renormalizable level \footnote{It is also true that 
one cannot write nonrenormalizable Yukawa interactions with  the SM 
Higgs doublet only. It is essential to include the Higgs doublets with nonzero 
$U(1)^{'}$ charges in order that one can write Yukawa couplings for the up-type
quarks in this model.}.
The number of additional Higgs doublets depends on the U(1)$^\prime$ charge 
assignment to the SM fermions,  especially the right-handed up-type quarks. 
In general, one must add three additional Higgs doublets with U(1)$^\prime$ 
charges $u_i$ [ see Refs.~\cite{u1models,u1models2} for more discussions ].
For the charge assignment $(u_i)=(0,0,1)$ we have two Higgs doublets
including the SM-like Higgs doublet, while for $(u_i)=(-1,0,1)$ three Higgs
doublets are required. The additional U(1)$^\prime$ must be broken in the end, 
so that we add a U(1)$^\prime$-charged singlet Higgs field $\Phi$ to the SM. 
Both the U(1)$^\prime$-charged Higgs doublet and the singlet $\Phi$ can give
the masses for the $Z^\prime$ boson and extra fermions if it has
a nonzero vacuum expectation value (VEV). After breaking of the electroweak 
and U(1)$^\prime$ symmetries, one can write down the Yukawa interactions 
in the mass basis. The Yukawa couplings depend on the masses 
of the involved quarks and the mixing angles between the Higgs fields 
and the right-handed up-type quarks, which rely on the U(1)$^\prime$ charge 
assignment~\cite{u1models,u1models2}. 
After all the Yukawa couplings would be proportional to the quark 
masses responsible for the interactions so that we could ignore 
the Yukawa couplings which are not related to the top quark. 

The number of relevant Higgs bosons participating in the top-quark pair 
production depends on the U(1)$^\prime$ charge assignment and mixing angles. 
The relevant Yukawa couplings for the top-quark pair production can be written as
\begin{equation}
V =  Y_{tu} \overline{u_L} t_R h+Y^H_{tu} \overline{u_L} t_R H 
+ i Y_{tu}^a \overline{u_L} t_R a + h.c.,
\end{equation}
where $h$ and $a$ are is the lightest neutral scalar and pseudoscalar 
Higgs bosons, and $H$ is the heavier (second lightest) neutral Higgs boson.
We assume that the Yukawa couplings of the other Higgs bosons 
are suppressed by the mixing angles \footnote{
This assumption is not compulsory, since all the Higgs bosons might 
participate in the top-quark pair production in principle. We will keep only
a few lightest (pseudo) scalar bosons in order to simplify the numerical analysis. 
}.

In the Ref.~\cite{u1models2}, the explicit expressions are
given in the $(u_i)=(0,0,1)$ case:
\begin{eqnarray} 
Y_{tu} & = & \frac{ 2m_t (g^u_R)_{ut} }{v \sin( 2 \beta)} 
\sin (\alpha-\beta) \cos \alpha_{\Phi} \ ,  
\\
Y^H_{tu} & = & - \frac{ 2m_t (g^u_R)_{ut} }{v \sin( 2 \beta)} 
\cos (\alpha-\beta) \cos \alpha_{\Phi} \ ,
\\
Y^a_{tu} & = & \frac{ 2m_t (g^u_R)_{ut} }{v \sin( 2 \beta)} \ .
\end{eqnarray} 
The $Y^{(a)}_{tt}$ couplings could be also large.  But in this case, 
the $s$-channel contribution of the Higgs bosons to the production of 
the top-quark pair would be suppressed  by the  $Y_{qq}^{(a)}$ couplings 
of light quarks,  which are proportional to $m_q$.

Finally, leptophobic and flavor-dependent chiral U(1)$^\prime$ 
models are anomalous. The gauge anomalies can be easily canceled 
by adding extra chiral fermions: for example, one extra generation 
and two SM gauge vector-like pairs~\cite{u1models2}. One of the fermions 
may be a good candidate for the dark matter and the Higgs boson 
could decay to two dark matters because of the mixing between the Higgs doublets
and the SM singlet field $\Phi$. If the branching ratio of the Higgs boson
to the dark matters is large, the stringent constraints from the Higgs boson 
search at the LHC could be relaxed \cite{u1models}, and the Higgs boson of mass 
around 200 GeV is still viable because it could decay into a pair of CDM's.

\section{Phenomenology \label{sec:pheno}}
\subsection{Generalities and Inputs}

In this section, we discuss phenomenology of our model described in the previous 
section. If new physics affects the top-quark pair production and could accommodate  
$A_\textrm{FB}^t$ at the Tevatron,  it must also be consistent with many other 
experimental measurements related with the top quark.
In our models, both the $Z^\prime$ and Higgs bosons $h$ and $a$  contribute
to the top-quark pair production through the $t$-channel exchange in 
the $u\bar{u} \to t \bar{t}$ process.  As we discussed in the previous section,
the $Z^\prime$ boson also  contributes to the top-quark pair production 
through the $s$-channel exchange, which was ignored in Ref.~\cite{Jung:2009jz}.

As two extreme cases, one can consider the cases where only the $Z^\prime$ boson
or Higgs boson $h$ contributes to the top-quark pair production.
Then, our models become close to the simple $Z^\prime$ model of Ref.~\cite{Jung:2009jz}
or the scalar-exchange model of Ref.~\cite{Babu:2011yw}. 
Unfortunately, these models cannot be compatible with the present upper bound 
on the same-sign top-quark pair production at the LHC in the parameter space 
which give rise to a moderate $A_\textrm{FB}^t$~\cite{u1models2,Aad:2012bb,Cao:2011ew}. 
In our chiral U(1)$^\prime$ models, the constraint from the same-sign top-quark
pair production could be relaxed because of the destructive interference
between the contribution from the $Z^\prime$ and those from Higgs bosons $h$ and $a$. 
In particular, the contribution of the pseudoscalar boson $a$ to the same-sign 
top-quark pair production is opposite to the other contributions.

In the two Higgs doublet model with the $U(1)^\prime$ assignments  
to the right-handed up-type quarks, $(u_i)=(0,0,1)$,  
the $s$-channel contribution of the $Z^\prime$ exchange 
to the partonic process  $u\bar{u}\to t\bar{t}$  is as strong as 
an $t$-channel contribution because of the relation 
$(g_R^u)_{ut}^2 = (g_R^u)_{uu} (g_R^u)_{tt}$~\cite{u1models2}.
In the multi-Higgs doublet models (mHDMs)  with other U(1)$^\prime$ charge 
assignments $(u_i)'$s to the right-handed up-type quarks, 
the $s$-channel contribution could be small.
In general, one can write $(g_R^u)_{uu} (g_R^u)_{tt} = \xi (g_R^u)_{ut}^2$,
where $\xi$ is a function of mixing angles and $0\leq |\xi| \leq O(1)$. 
In the case of $m_{Z^\prime} \geq 2 m_t$, a resonance around the $Z^\prime$ 
mass for nonzero $\xi$ would be observed in the $t\bar{t}$ invariant mass 
distribution.  However, such a resonance has not been observed so far in the experiments~\cite{cmsmtt,atlasmtt}.  This would restrict the $Z^\prime$ mass to be 
much smaller than $2 m_t$ for nonzero $\xi$.

The cross sections for the top-quark pair production at the Tevatron are 
$\sigma(t\bar{t})=(7.5\pm 0.48)$ pb at CDF~\cite{cdfttbar} and 
$\sigma(t\bar{t})= (7.56^{+0.63}_{-0.56})$ pb at D0~\cite{d0ttbar}, 
respectively. At the LHC, the cross sections for the top-quark pair production
are $\sigma(t\bar{t})=(165.8\pm 13.3)$ pb at CMS~\cite{cmsttbar} and 
$\sigma(t\bar{t})=(177\pm 11)$ pb at ATLAS~\cite{atlasttbar}, respectively.
In this work, we require that the cross section for the top-quark pair production
is in agreement with the CDF result in the $1\sigma$ level, which has
the least uncertainty. Another reason to use the Tevatron result for the check
of our model is that the top-quark pair production at the Tevatron is more
sensitive to new physics in the $u\bar{u}\to t\bar{t}$ process than at the LHC.

In the SM, the top quark dominantly decays into $W+b$. In our models, there are
several flavor-changing vertices $u_R$-$t_R$-$Z^\prime$, $u_R$-$t_L$-$h$, 
and $u_R$-$t_L$-$a$. If the $Z^\prime$ or Higgs bosons are lighter than the top 
quark, it could be dangerous because the branching ratio of the top quark
to $W+b$ is significantly altered. We assume that the pseudoscalar Higgs boson 
$a$ is heavier than the top quark and the branching ratio of 
the exotic decay of the top quark such as $t \rightarrow Z^\prime u , h u$ 
is less than 5\%. 
We find that  the exotic decay mode of the top quark can be suppressed, 
if we choose $\alpha_x \equiv ((g^\prime g_R^u)_{ut} )^2/(4\pi) \lesssim 0.012$ 
for $m_{Z^\prime} = 145$ GeV and $Y_{tu} \lesssim 0.5$ for $m_h=125$ GeV. 

Furthermore,  such large FCNCs could generate the same-sign top-quark pair 
production  through the $t$-channel diagram in the $uu \to tt$ process,  
which is forbidden within the SM~\cite{Cao:2011ew}. 
The CMS Collaboration announced the upper bound on the cross section
for the same-sign top-quark pair production: 
$\sigma^{tt} < 17$ pb at 95\% CL. with a luminosity of
35 pb$^{-1}$~\cite{Chatrchyan:2011dk},
while the limit on the cross sections at ATLAS with a luminosity of 
$1.04$ fb$^{-1}$ are
$\sigma^{tt} < 2$ pb at 95\% CL. by using an optimized event selection 
for the $Z^\prime$ model and
$\sigma^{tt} < 4$ pb at 95\% CL. by using more inclusive selection,
respectively~\cite{Aad:2012bb}. We use the latter limit in this work.

In numerical analysis, we take the top-quark mass to be $m_t=173$ GeV.
For a parton distribution function we use CTEQ6m with the renormalization
and the factorization scale equal to $\mu=m_t$~\cite{cteq}. 
In order to take into account the QCD radiative correction which is unknown 
as of now for the model under consideration, we use the $K$ factor obtained 
in the perturbative QCD calculations: namely, $K=1.3$ for the Tevatron and 
$K=1.7$ for the LHC by assuming the same $K$ factor in the new physics model. 
The center-of-momentum energy $\sqrt{s}$ is $1.96$ TeV at the Tevatron and 
$7$ TeV at the LHC, respectively.
In the previous works~\cite{u1models,u1models2,u1models3}, we did not consider 
the SM NLO contribution to $A_\textrm{FB}^t$, but in this work
we take into account its contribution to $A_\textrm{FB}^t$ by using
the approximated formula $A_\textrm{FB}^t \simeq A_\textrm{FB}^{t,\textrm{SM}}
+\delta A_\textrm{FB}^{t}/K$, where the first term denotes
$A_\textrm{FB}^t$ at the SM NLO and the second one corresponds to
the contribution from the new physics. 
We also use the approximated formula $A_C^y \simeq A_C^{y,\textrm{SM}}
+\delta A_C^y/K$.

\begin{figure}[!t]
\begin{center}
\epsfig{file=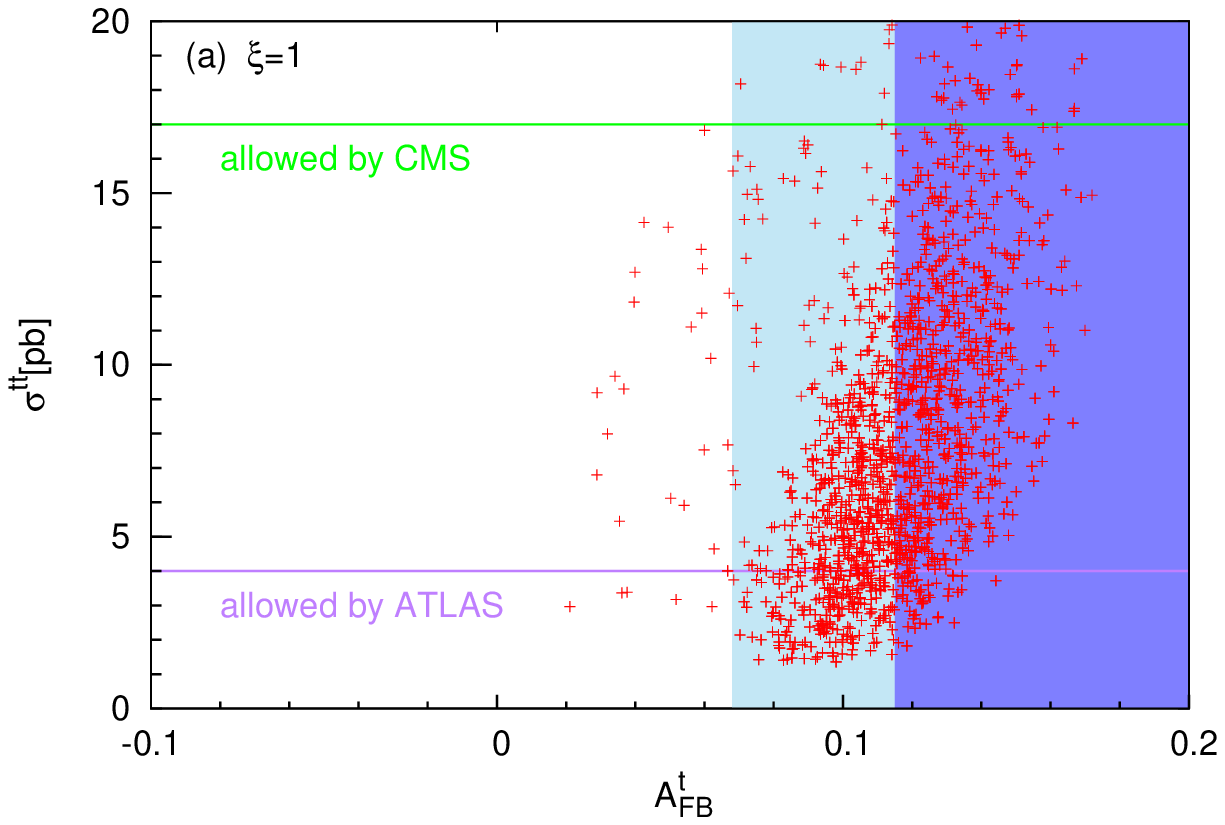,width=0.45\textwidth}
\epsfig{file=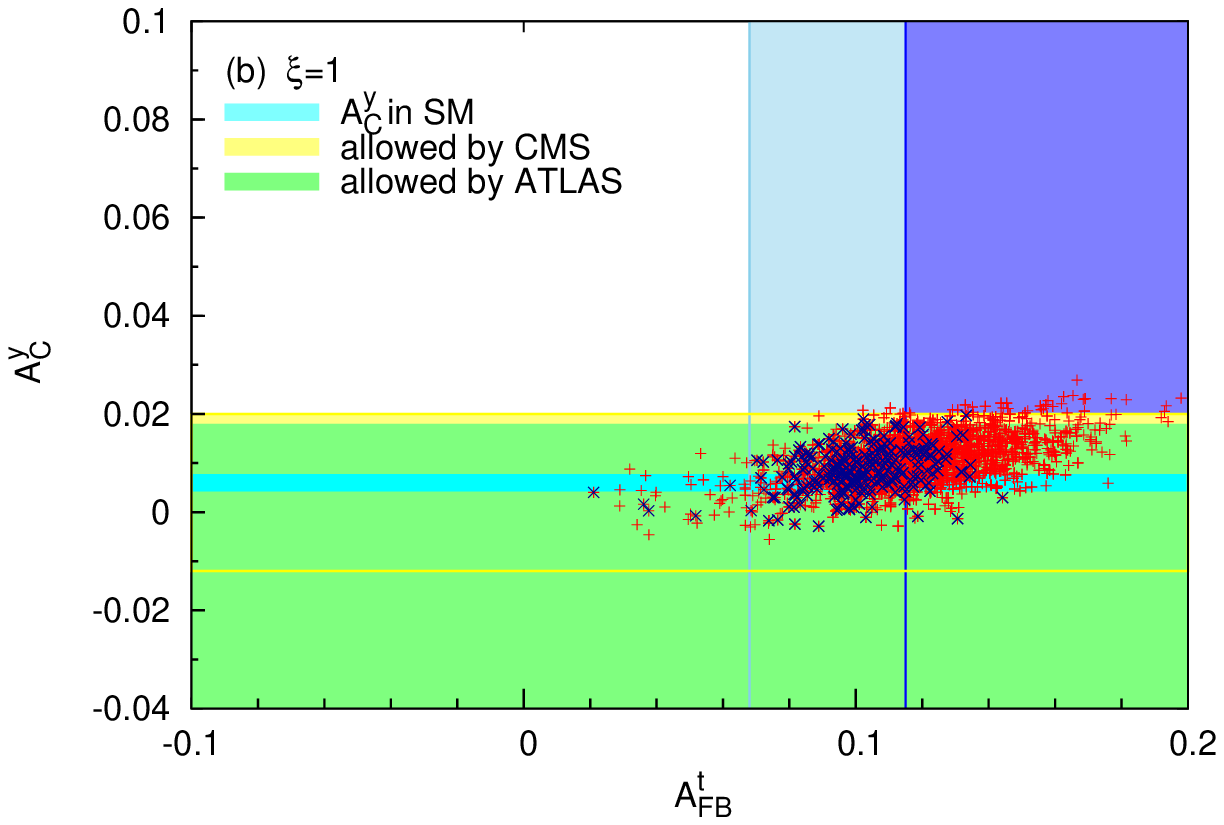,width=0.45\textwidth}
\caption{\label{fig:lightzp}%
The scattered plots for (a) $A_\textrm{FB}^t$ at the Tevatron and
$\sigma^{tt}$ at the LHC in unit of pb, and (b) $A_\textrm{FB}^t$ at the 
Tevatron and $A_C^y$ at the LHC for $m_{Z^\prime}=145$ GeV and $\xi=1$.
In (b), the blue points satisfy the upper bound on the same sign top pair 
production from ATLAS: $\sigma^{tt} < 4$ pb.
}
\end{center}
\end{figure}

\begin{figure}[!t]
\begin{center}
\epsfig{file=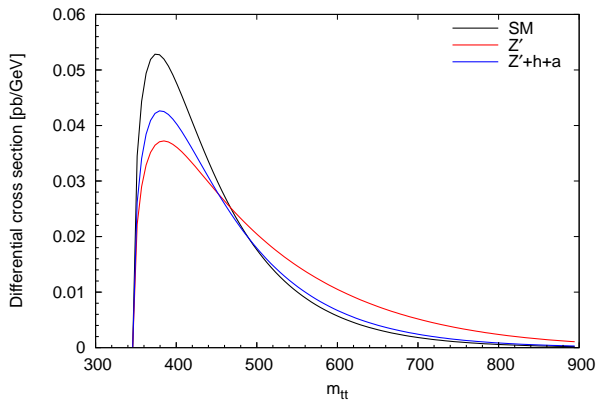,width=0.45\textwidth}
\caption{\label{fig:mtt}%
The invariant mass distribution of the top-quark pair at the Tevatron
in the SM, $Z^\prime$ model, and chiral U(1)$^\prime$ model.
}
\end{center}
\end{figure}

\subsection{ $m_{Z^{'}}=145$ GeV cases
\label{sec:mzp} }

In this subsection, we discuss the light $Z^\prime$ cases 
with $m_{Z^\prime} = 145$ GeV. 
The scalar Higgs boson $h$ is chosen to be heavier than the top quark
in order to suppress the branching ratio of the exotic decay of the top quark.
In the two Higgs doublet model (2HDM), $\xi$ is fixed to be 1 so that the $s$-channel 
exchange of the $Z^\prime$ boson is as strong as the $t$-channel exchange.
On the other hand, in the mHDM cases, $|\xi|$ would be between 0 and $O(1)$ 
depending on the mixing angles between scalar bosons.
In the case of $\xi \simeq 0$, the $s$-channel contribution of $Z^\prime$ 
would be negligible.

In this work, we consider two cases, $\xi=1$ and $\xi=0$ for illustration.

\subsubsection{$\xi=1$ case
\label{sec:xi1-1}}

Here, we consider the case with $\xi=1$, so that one has the relation 
$(g_R^u)_{uu} (g_R^u)_{tt} = \xi (g_R^u)_{ut}^2$.  
Then the $s$-channel $Z^{'}$ exchange contribution is compatible 
with the $t$-channel $Z'$ exchange contribution, and both should be 
kept in the phenomenological analysis. 

The other model parameters are chosen in the following ranges:
$180~\textrm{GeV} \leq m_h, m_a \leq 1$ TeV, 
$0.005 \leq \alpha_x \leq 0.012$, 
and $0.5 \leq Y_{tu}, Y_{tu}^a \leq 1.5$ with the condition 
$Y_{tu}\leq Y_{tu}^a$ [ see Eq.s~(5)-(7) ]. 
In Fig.~\ref{fig:lightzp} (a), we show the scattered plot for $A_\textrm{FB}^t$
at the Tevatron and $\sigma^{tt}$ at the LHC at 7 TeV in unit of pb for 
$m_{Z^\prime}=145$ GeV. 
The red points satisfy the cross section for the top-quark pair 
production at the Tevatron in the 1$\sigma$ level. 
The regions below horizontal lines are the allowed regions from the limits on 
$\sigma^{tt}$ at CMS and ATLAS, respectively. 
As discussed in Ref.~\cite{Aad:2012bb}, the $t$-channel exchange models 
with a $Z^\prime$ or scalar boson only are disfavored by the constraint on 
$\sigma^{tt}$ at the LHC.  But in our model, this strong constraint could be
relaxed due to the destructive interferences between the $Z^\prime$ and Higgs
boson $t$-channel exchange diagrams. 

This is one of important lessons in our study for the flavor-dependent chiral 
U(1)$^\prime$ model. It is often argued that a certain model is excluded or 
disfavored from experiments by assuming that only one coupling is dominant 
and ignoring other contributions. 
However, if a complete model with all the necessary ingredients is considered,
the stringent constraint from experiments might be relaxed.

In Fig.~\ref{fig:lightzp} (a), the blue and skyblue bands are consistent with
$A_\textrm{FB}^t$ in lepton+jets channels at CDF in the 1$\sigma$ and
2$\sigma$ levels, respectively. We note that there exists a favored region
even if we use the most stringent constraint on $\sigma^{tt}$ at ATLAS.
Our model with the light $Z^{'}$ boson could be disfavored, if the experimental 
upper bound on the same-sign top-quark pair production cross section 
becomes below $\sim 1$ pb at the LHC in the near future.

In Fig.~\ref{fig:lightzp} (b), we show the scattered plot for $A_\textrm{FB}^t$
at the Tevatron and $A_C^y$ at the LHC at 7 TeV for $m_{Z^\prime}=145$ GeV. 
The yellow and green regions are experimental bounds on $A_C^y$ in the 
1$\sigma$ level at CMS and at ATLAS, respectively. The horizontal cyan band is
the SM prediction for the charge asymmetry $A_C^y$. The blue and skyblue bands
are same as in Fig.~\ref{fig:lightzp} (a). 
The red dots in Fig.~\ref{fig:lightzp} (b) satisfy 
the $t\bar{t}$ production cross section at the Tevatron within 1$\sigma$ 
and the blue dots satisfy the experimental limit  $\sigma^{tt}<4$ pb
on the cross section for the same-sign top-quark pair production at the LHC
as well as the $t\bar{t}$ production cross section at the Tevatron
within 1$\sigma$. 
The $Z^\prime$ or $W^\prime$ exchange model in the $t$-channel predicts
a large positive $A_C^y$, which might be inconsistent with the current data
at the LHC~\cite{atlasacy}. The light $Z^\prime$ case in our model is
in good agreement with the data for $A_C^y$ at the LHC  in some parameter regions 
as shown in Fig.~\ref{fig:lightzp} (b) due to the additional contributions from the
neutral Higgs bosons $h$ and $a$.

The invariant mass distribution of the top-quark pair (especially in the large 
invariant mass region) could be a good discriminator of the models for 
$A_\textrm{FB}^t$~\cite{Jung:2009jz}.
In Fig.~\ref{fig:mtt}, we show the invariant mass distribution of the top quark 
pair produced at the Tevatron. The black curve is the SM case
at leading order (LO). The red curve corresponds to the original $Z^\prime$ 
model  (without neutral Higgs bosons) with a large off-diagonal coupling for
$m_{Z^\prime}=145$ GeV and $\alpha_x= 0.029$~\cite{Jung:2009jz}. 
In this case, the model overestimates (underestimates) the SM predictions 
in the large (small) invariant  mass region.
This is a typical feature of the model with a large $t$-channel contribution 
to the $q\bar{q} \to t\bar{t}$. Finally, the blue curve is 
the chiral U(1)$^\prime$ model (with the contributions 
of $Z^{'}$, $h$ and $a$ all included) with the following parameters: 
$m_{Z^\prime}=145$ GeV, $m_{h}=180$ GeV, $m_{a}=250$ GeV, 
$\alpha_x=0.005$, $Y_{tu}=1$, and $Y_{tu}^a=1.1$.
The general feature is similar to the $Z^\prime$ model, but
the prediction of the chiral U(1)$^{\prime}$ model becomes much closer 
to the LO SM prediction because of the destructive interferences
between the contributions from the $Z^\prime$ and Higgs bosons.
This is another benefit of our model,  since the current measurement of
 the $t\bar{t}$ invariant mass distribution is not much deviated from the 
SM prediction. For more detailed comparison,  one must include the NLO 
predictions in the SM and each new physics model, which is not available 
yet in the literature and also beyond the scope of this paper.
However, we could conclude that the large deviation of the $t\bar{t}$ 
invariant mass distribution in the original $Z^\prime$ model can be 
significantly improved by including the contributions of Higgs bosons 
$h$ and $a$.

\begin{figure}[!th]
\begin{center}
\epsfig{file=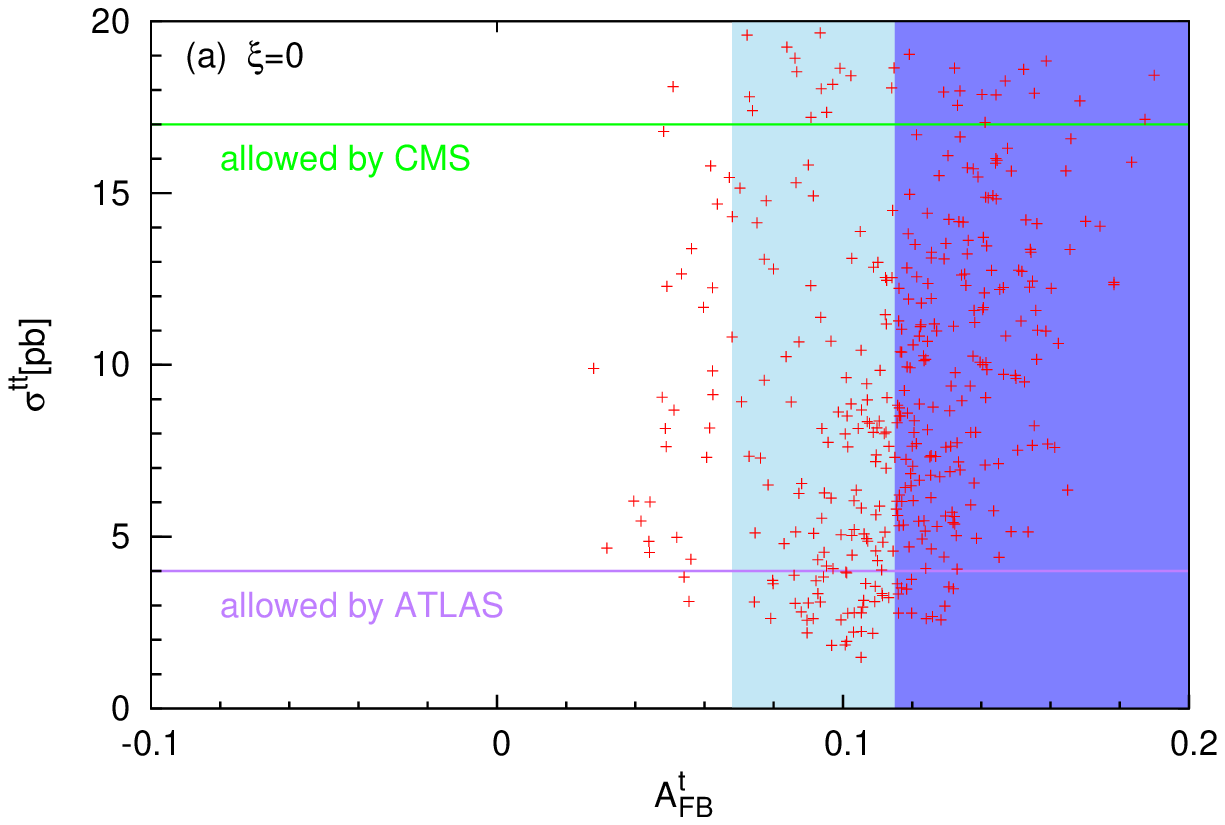,width=0.45\textwidth}
\epsfig{file=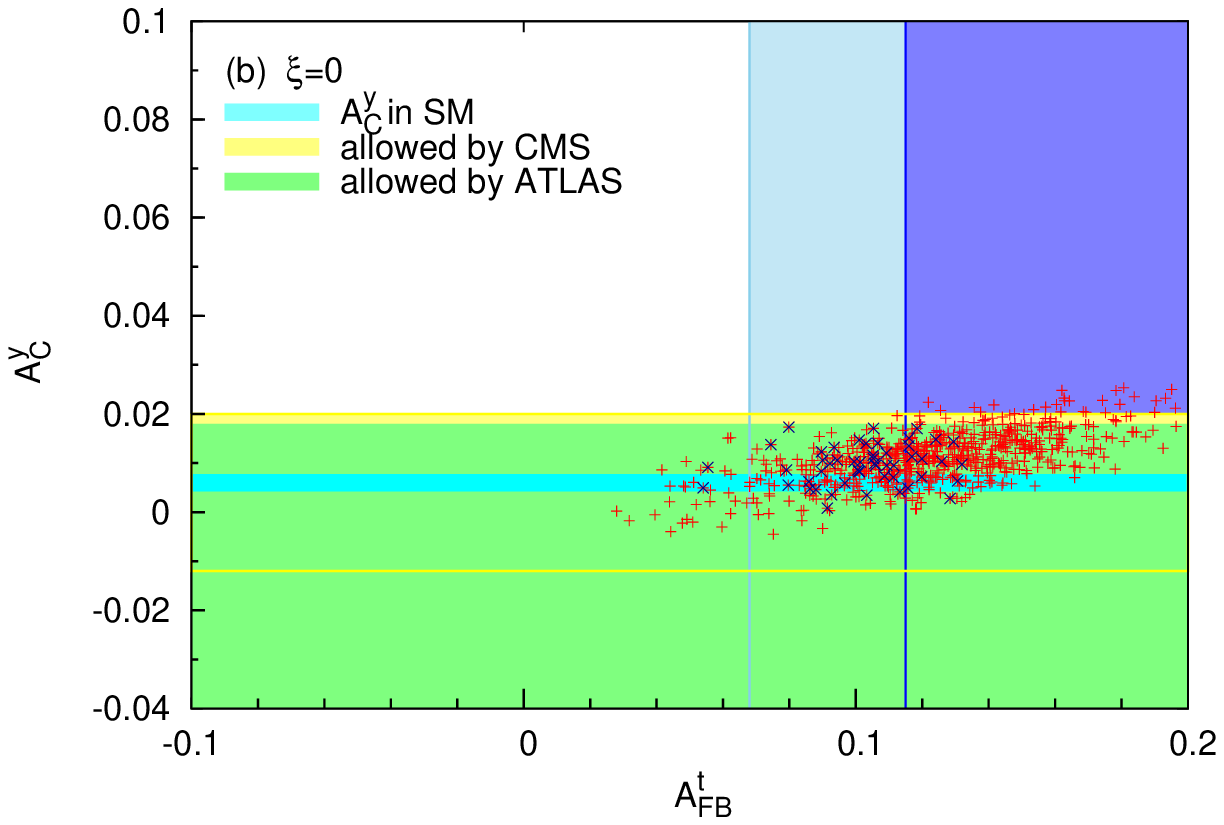,width=0.45\textwidth}
\caption{\label{fig:lightzp0}%
The scattered plots for (a) $A_\textrm{FB}^t$ at the Tevatron and
$\sigma^{tt}$ at the LHC in unit of pb, and (b) $A_\textrm{FB}^t$ at the 
Tevatron and $A_C^y$ at the LHC for $m_{Z^\prime}=145$ GeV and $\xi=0$.
}
\end{center}
\end{figure}

\subsubsection{$\xi=0$ case
\label{sec:xi0-1}%
}

Now, let us discuss another extreme case with $\xi=0$: 
namely, $(g_R^u)_{uu} (g_R^u)_{tt} = 0$. 
Then, the $Z^\prime$ boson contributes to the top-quark pair production
through only its $t$-channel exchange.  And there would be no strong 
constraints from dijet or $t\bar{t}$  resonance searches.
We vary other model parameters in the same ranges as in the 
$\xi=1$ case  of section~\ref{sec:xi1-1}.
In Fig.~\ref{fig:lightzp0}, we present the scattered plots (a) for 
$A_\textrm{FB}^t$ at the Tevatron and $\sigma^{tt}$ at the LHC 
in unit of pb and
(b) for $A_\textrm{FB}^t$ at the Tevatron and $A_C^y$ at the LHC, 
where we use the same legends as in Fig.~\ref{fig:lightzp}. 
The general feature is basically the same as in Fig.~\ref{fig:lightzp}. 
We find that there is a parameter space where our predictions are in good 
agreement with the current experimental constraints in case of $\xi=0$.
This would also imply that there would be some parameter regions satisfying 
the empirical data in the range  $0\le \xi \le 1$.

\begin{figure}[!t]
\begin{center}
\epsfig{file=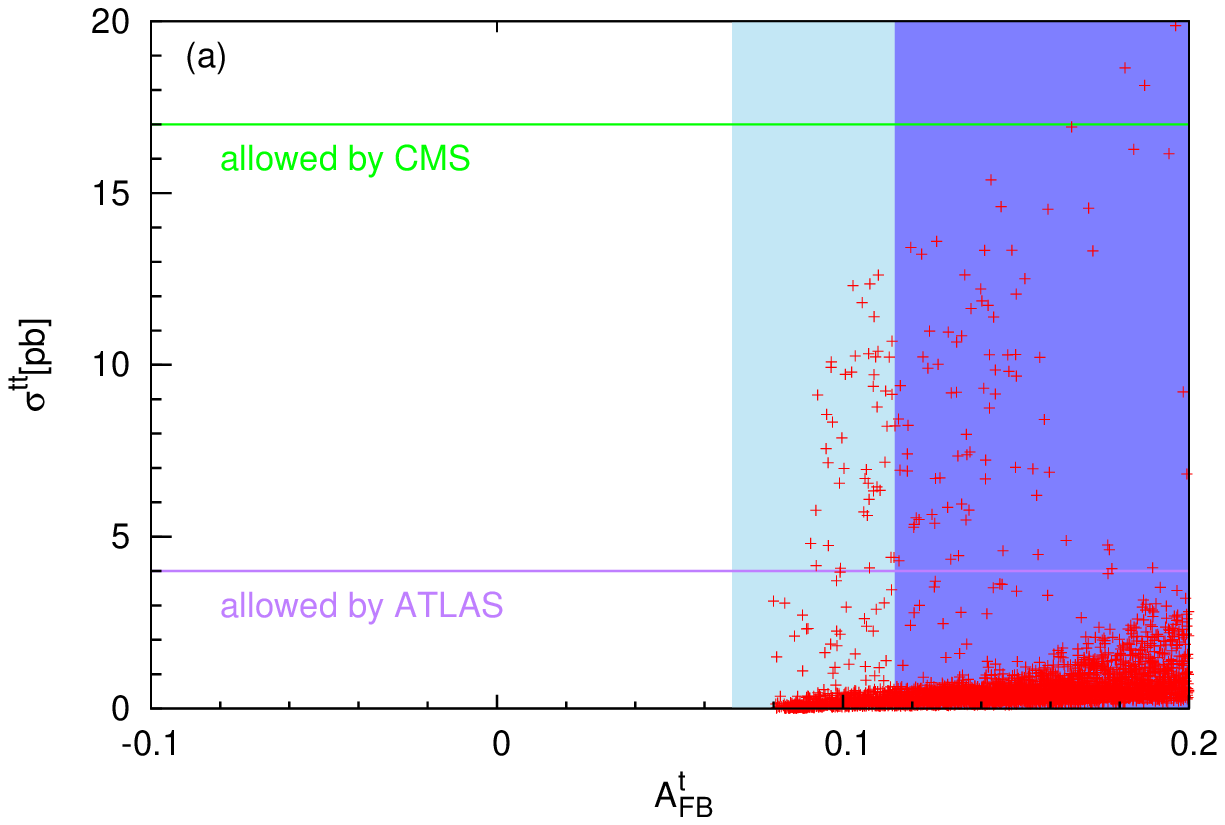,width=0.45\textwidth}
\epsfig{file=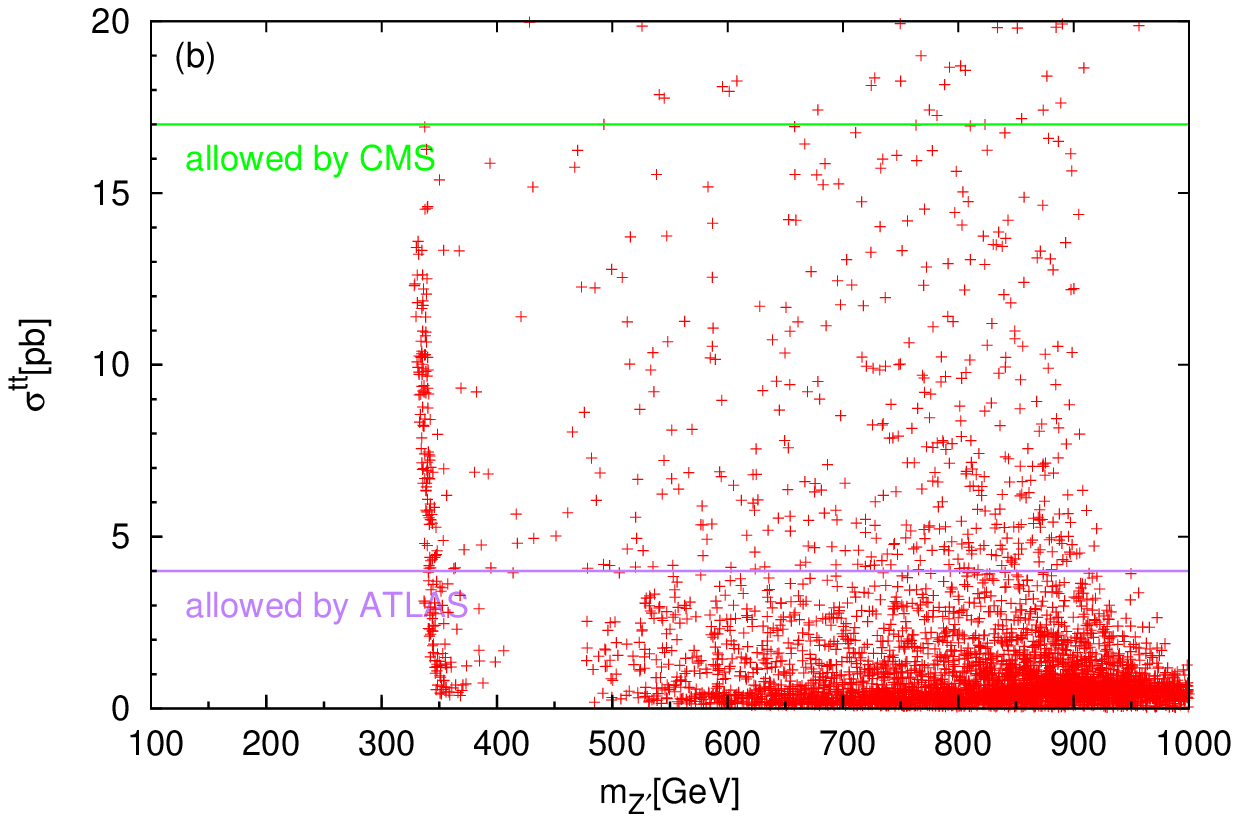,width=0.45\textwidth}
\caption{\label{fig:higgs1}%
The scattered plots for (a) $A_\textrm{FB}^t$ at the Tevatron and
$\sigma^{tt}$ at the LHC in unit of pb, and (b) $m_{Z^\prime}$ in unit of GeV 
and $\sigma^{tt}$ at the LHC in unit of pb for $m_h=125$ GeV.
}
\end{center}
\end{figure}

\subsection{$m_h=125$ GeV within 2HDM ($\xi=1$)
\label{sec:xi1-3}%
}
In the previous works~\cite{u1models,u1models2,u1models3}, 
only the relatively light $Z^\prime$ case was considered, since we were 
also interested in accounting for the $Wjj$  excess at CDF in the same model. 
Because the CDF $Wjj$ excess was not confirmed by the D0 Collaboration, 
the motivation for $m_{Z^\prime}\simeq 145$ GeV becomes weaker.
On the other hand, both ATLAS and CMS announced discovery of new boson of 
mass around 125 GeV ~\cite{ATLAS:2012ae,Chatrchyan:2012tx}, whose properties 
are quite similar to those of the SM Higgs boson within experimental uncertainties.  

Therefore in this subsection, we consider $m_h = 125$GeV motivated by the recent data ~\cite{ATLAS:2012ae,Chatrchyan:2012tx}, assuming the $Z^\prime$  mass is set free,  
Other parameters are chosen in the following ranges:
$180~\textrm{GeV} \leq m_{Z^\prime} \leq 1.5$ TeV, 
$180~\textrm{GeV} \leq m_a \leq 1$ TeV, $0.005 \leq \alpha_x \leq 0.025$, 
$0.1 \leq Y_{tu} \leq 0.5$, and
$0.1 \leq Y_{tu}^a \leq 1.5$ 
with the condition $Y_{tu}\leq Y_{tu}^a$ and $\xi=1$.
We note that the Yukawa coupling $Y_{tu}$ is chosen to be less than $0.5$ 
in order to satisfy the condition 
$\textrm{Br}(t\to \textrm{non-SM state}) \lesssim 5\%$.
In Fig.~\ref{fig:higgs1} (a), we show the scattered plot 
for $A_\textrm{FB}^t$ at the Tevatron and $\sigma^{tt}$ in unit of pb 
at the LHC for the lightest Higgs boson mass $m_h=125$ GeV. 
As in Fig.~\ref{fig:lightzp} (a), the red points satisfy the cross section
for the $t\bar{t}$ production at the Tevatron within 1$\sigma$.
Many points in the right-bottom seem to be in good agreement 
with the constraints from the top FB asymmetry at the Tevatron 
and the same-sign top pair production at the LHC. However, we find that
the $Z^\prime$ mass for those points are in the range of 
$350~\textrm{GeV} \lesssim m_{Z^\prime} \lesssim 1.2$ TeV as shown
in Fig.~\ref{fig:higgs1} (b). 
As we have discussed earlier, the $s$-channel diagram through the $Z^\prime$ 
exchange contributes to the $u\bar{u}\to t\bar{t}$ process  for $\xi=1$. 
This implies that in the $t\bar{t}$ invariant mass distribution
there appears a sharp peak around the $Z^\prime$ boson mass, which has not been
observed in experiments~\cite{cmsmtt,atlasmtt}. 

If we choose $\xi=0$ or the decay width of the $Z^\prime$ boson is sufficiently 
large,  then the resonance peak may not be observed in experiments  for 
$m_{Z^\prime} > 2 m_t$.  In our $U(1)^{'}$ model, the latter case cannot be realized. 
However, the former case might be possible for a certain mixing angles in the mHDMs. 
Therefore, we searched for the parameter space which is consistent with all the experiments  
by setting $\xi=0$.  However, we found no favored region in this case too. 
In particular, $A_\textrm{FB}^t$ could not be accommodated with the CDF data 
in the 1$\sigma$ level.  However, if we search with more relaxed experimental constraints,
for example, the 2$\sigma$ level for $A_\textrm{FB}^t$, we find that
there exist the favored region consistent with all the experiments.

\begin{figure}[!t]
\begin{center}
\epsfig{file=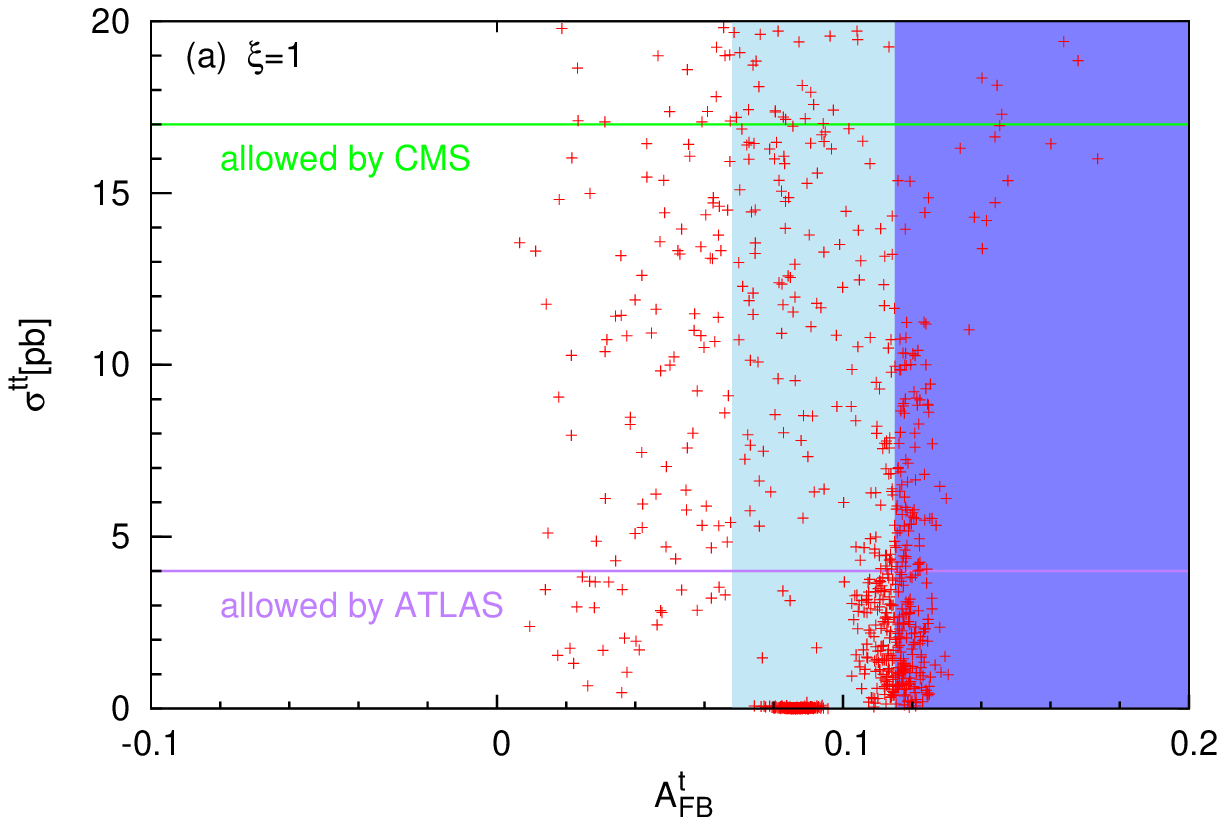,width=0.45\textwidth}
\epsfig{file=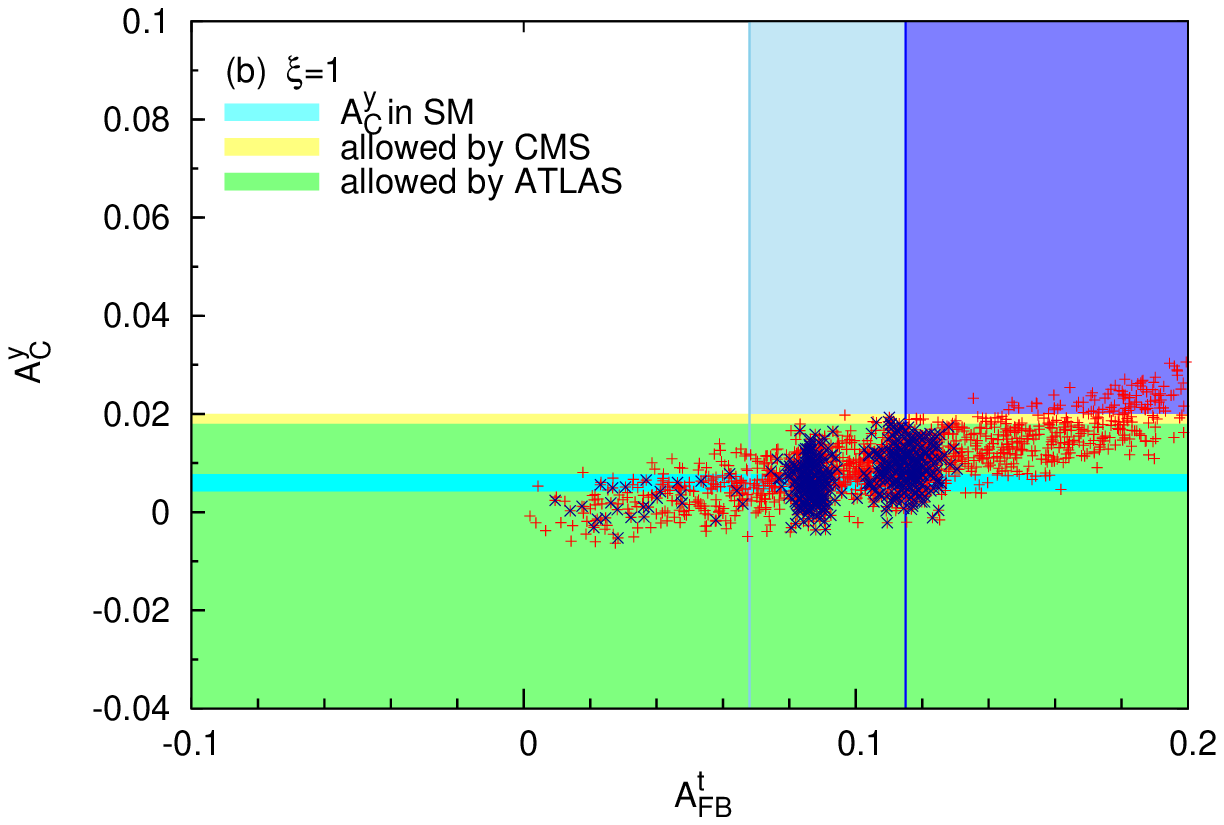,width=0.45\textwidth}
\caption{\label{fig:higgs2}%
The scattered plots for (a) $A_\textrm{FB}^t$ at the Tevatron and
$\sigma^{tt}$ at the LHC in unit of pb, and (b) $A_\textrm{FB}^t$ at the 
Tevatron and $A_C^y$ at the LHC for $m_h=125$ GeV and $\xi=1$,
where the contribution of the second lightest Higgs boson $H$ 
is included.
}
\end{center}
\end{figure}

\subsection{More Higgs bosons included
\label{sec:more}%
}
Up to now, we have kept only the lightest scalar ($h$) and pseudoscalar 
($a$) Higgs bosons, assuming other heavier Higgs bosons decouple from 
top physics because either they are very heavy or they  have small Yukawa couplings. 
In this subsection, we will relax this assumption, and  include the second lightest 
Higgs boson $H$ in the analysis. For the completeness, one must consider
all the Higgs bosons as well as the extra fermions, but it would be 
quite complicated and time-consuming since there are too many new parameters 
involved.  We will be content with a simplified discussion of Higgs sector.  
In the following sections,  we consider two cases $\xi=1$ and $\xi=0$ for illustration.

\subsubsection{$\xi=1$ case
\label{sec:xi1-2}%
}
Let us first consider the $\xi=1$ case.  If the second lightest scalar Higgs boson 
$H$ has a flavor-changing coupling to the top quark like the lightest scalar
Higgs boson $h$,  there are four new particles $Z^\prime$, $h$, $H$, and $a$ 
that contribute to the top-quark pair production.
We take the lightest Higgs boson mass to be $m_h = 125$ GeV like 
in the previous case. The other parameters are taken to be 
in the following ranges:
$160~\textrm{GeV} \leq m_{Z^\prime} \leq 300$ GeV,
$180~\textrm{GeV} \leq m_H, m_a \leq 1$ TeV, 
$0 \leq \alpha_x \leq 0.025$, 
$0 \leq Y_{tu} \leq 0.5$, and
$0 \leq Y_{tu}^H, Y_{tu}^a \leq 1.5$, 
where $m_H$ is the mass of the second lightest scalar Higgs boson
and $Y_{tu}^H$ is its Yukawa coupling to the $u$-$t$-$H$ vertex.
Note that the range of the $Z^\prime$ boson mass is chosen 
to avoid the constraints from the top quark decay and $t\bar{t}$ 
invariant mass distribution.

Figure~\ref{fig:higgs2} (a) shows the scattered plot for 
$A_\textrm{FB}^t$ at the Tevatron and $\sigma^{tt}$ at the LHC  in unit of pb. 
The red points satisfy the cross section for the top-quark pair production 
at the Tevatron in the 1$\sigma$ level.
The horizontal lines are upper limits on the same-sign top-quark pair production
at CMS and ATLAS, respectively.
The blue and skyblue regions are consistent with $A_\textrm{FB}^t$
in the lepton+jets channel at CDF within 1$\sigma$ and 2$\sigma$, respectively.
We find that there exist some parameter regions which
are in agreement with $A_\textrm{FB}^t$ within 1$\sigma$ and 
the upper limit on $\sigma^{tt}$ at ATLAS.
Thus the light Higgs boson with $m_h=125$ GeV in our model could pass all 
the experimental constraints if the heavier scalar Higgs boson $H$ contributes
to the top-quark pair production.

In Fig.~\ref{fig:higgs2} (b), we show the scattered plot for $A_\textrm{FB}^t$
at the Tevatron and $A_C^y$ at the LHC. 
Each region on the figure denotes the same 
experimental constraint as on Fig.~\ref{fig:lightzp} (b).
The red points in Fig.~\ref{fig:higgs2} (b) satisfy 
the $t\bar{t}$ production cross section at the Tevatron within 1$\sigma$ 
and the blue points satisfy the experimental limit  $\sigma^{tt}<4$ pb
on the cross section for the same-sign top-quark pair production at ATLAS
as well as the $t\bar{t}$ production cross section at the Tevatron
within 1$\sigma$. 
The light Higgs boson case with $m_h=125$ GeV could be
in good agreement with the data for $A_\textrm{FB}^t$ at the Tevatron and
$A_C^y$ at the LHC as shown in Fig.~\ref{fig:higgs2} (b) if more Higgs bosons 
are included.   Furthermore, it is amusing that in this case the same-sign top-quark 
pair production at the LHC  could be less than 1 pb as shown in Fig.~\ref{fig:higgs2} (a).

\begin{figure}[!th]
\begin{center}
\epsfig{file=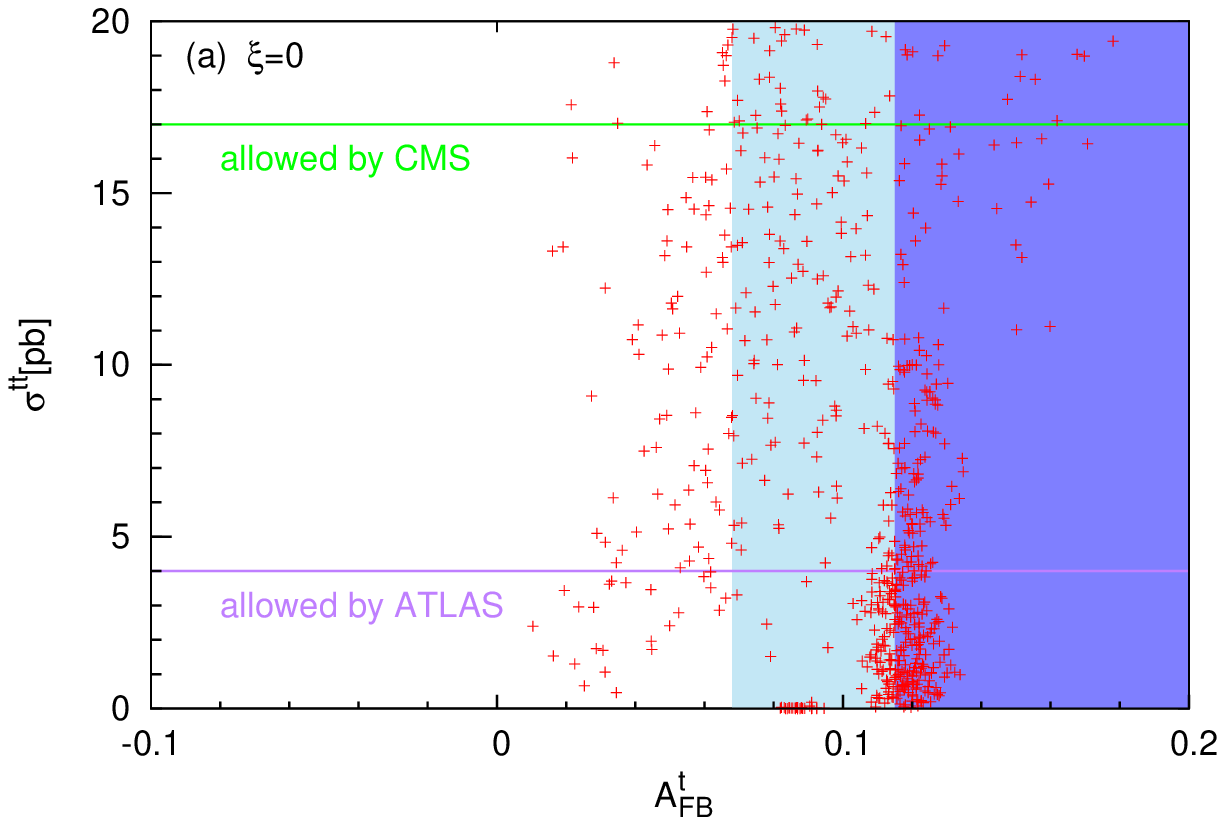,width=0.45\textwidth}
\epsfig{file=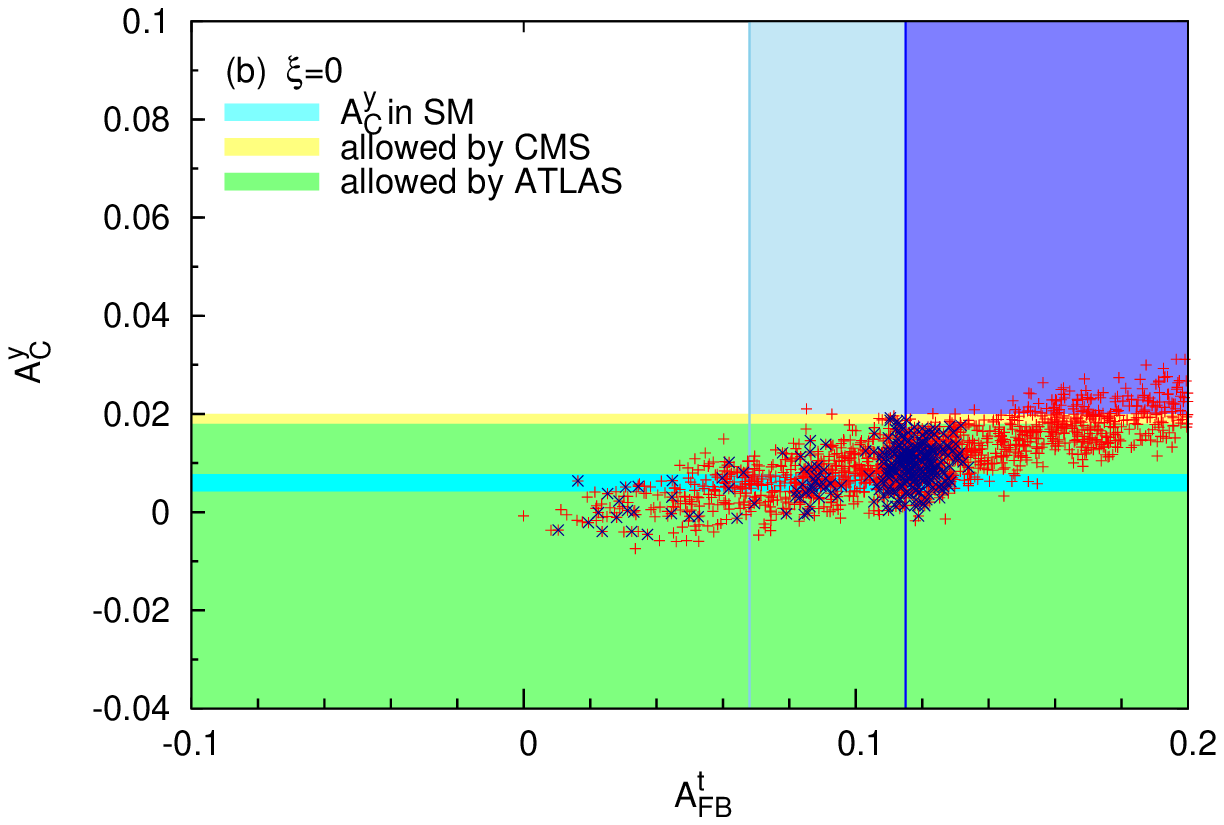,width=0.45\textwidth}
\caption{\label{fig:higgs2-2}%
The scattered plots for (a) $A_\textrm{FB}^t$ at the Tevatron and
$\sigma^{tt}$ at the LHC in unit of pb, and (b) $A_\textrm{FB}^t$ at the 
Tevatron and $A_C^y$ at the LHC for $m_h=125$ GeV and $\xi=0$,
where the contribution of the second lightest Higgs boson $H$ is included.
}
\end{center}
\end{figure}

\subsubsection{$\xi=0$ case
\label{sec:xi0-2}%
}

In this section, we consider the $\xi=0$ case, where the $s$-channel
contribution of the $Z^\prime$ boson to the top-quark pair production
is negligible. The other model  parameters are varied in the same ranges as 
in the $\xi=1$ case of Sec.~\ref{sec:xi1-2}. In Fig.~\ref{fig:higgs2-2}, we show
the scattered plots (a) for $A_\textrm{FB}^t$ at the Tevatron and 
$\sigma^{tt}$ at the LHC in unit of pb and (b) for $A_\textrm{FB}^t$
at the Tevatron and $A_C^y$ at the LHC, where all the legends on the figure
are the same as those on Fig.~\ref{fig:higgs2}.
As in the case of $\xi=1$, we  find some parameter regions satisfying
all the experimental constraints. Like the light $Z^\prime$ case,
one would find the allowed regions in the whole range of $0\le \xi \le 1$.

\subsection{Summary}

In this section, we examined three different cases of  chiral U(1)$^{\prime}$ 
models proposed by the present authors~\cite{u1models,u1models2,u1models3}:
(i) the light $Z^\prime$ model with $m_{Z^\prime} =145$ GeV and  
$h$ and $a$ being heavier than the top quark, 
(ii) the light scalar Higgs model with $m_h = 125$ GeV and 
$Z^\prime$ and $a$ being heavier than the top quark,
and (iii) the light scalar Higgs model with $Z^\prime$, $H$, and $a$, 
where $m_h = 125$ GeV, $m_{Z^\prime} \ge 160$ GeV and 
$m_{H,a}\ge m_t$.

In the first and third cases we can find the parameter regions
which are in good agreement with all the experimental constraints
in the 1$\sigma$ level, but in the second case the accommodation is
possible only in the 2$\sigma$ level. Except the above cases, there can 
exist other cases which might be consistent with experiments.
For instance, one can consider the light $Z^\prime$ and light Higgs model.
Suppose that $m_{Z^\prime}=145$ GeV and $m_h=125$ GeV, but
the lightest Higgs boson weakly couples to the top quark.
Then, we can apply the case (i) to this model by assuming 
that the lightest Higgs boson has a negligible off-diagonal Yukawa coupling
and the second Higgs boson has a large off-diagonal Yukawa coupling.

\section{Conclusions \label{sec:con}}
$A_\textrm{FB}^t$ reported by CDF and D0 Collaborations has been one 
of the hottest issues in particle physics phenomenology during last four years.  
Assuming that the observed deviation in $A_\textrm{FB}^t$ might be a signal of 
new physics,  a number of models have been proposed to explain the observed 
$A_\textrm{FB}^t$.  Some models were already disfavored by experiments,  
and other models might be verified at the LHC soon.

In the Refs.~\cite{u1models,u1models2,u1models3},  the present authors have 
proposed complete realistic models with U(1)$^{\prime}$ flavor symmetry, 
where only right-handed up-type quarks are charged flavor-dependently, 
in order to avoid the strong constraints on FCNC. Then, realistic Yukawa 
interactions were constructed by adding extra Higgs doublets which are charged 
under U(1)$^{\prime}$ flavor symmetry. Then, it was found that there arise 
new additional FCNC Yukawa couplings involving the light (pseudo)scalars.
Surprisingly, the interference between the extra gauge boson and scalar bosons
can relax the bound from the same-sign top process at LHC, and
we found that the light $Z'$ scenario can be revived. 

In this work, we reexamined the U(1)$^{\prime}$ models, including the 
updated results on the same-sign top-quark pair productions at the LHC, 
and investigated the top charge asymmetry at LHC. We found that there are 
parameter regions in the light $Z^{'}$ model with scalar 
and pseudoscalar Higgs bosons, which are consistent with empirical data,
for example, for $m_{Z^\prime}=145$ GeV, $m_h \sim 200$ GeV, 
and $m_a\sim 250$ GeV. However, if the same-sign top-quark pair production rate 
turns out to be smaller than $\sim 1$ pb,  this model with the U(1)$^{\prime}$  
and light $Z^{'}$ will be likely excluded.

Furthermore, the recent Higgs search at the LHC, which has a tentative excess 
in about $125$ GeV mass region, also constrains our scenario strongly. 
We also analyzed the light Higgs boson case. 
In the case of the light Higgs boson case with $Z^{\prime}$, $H$, and $a$, 
where $m_h = 125$ GeV, $m_{Z^\prime}\ge 160$ GeV and $m_{H,a}\ge m_t$,
there exist parameter spaces which can be in good agreement with
all the empirical data within 1$\sigma$.  Even though the upper bound on
the same-sign top-quark pair production rate at the LHC becomes less than 1 pb,
we can find some parameter regions consistent with the experimental data.
However, in the case that only the lightest Higgs boson $h$ ($m_h= 125$ GeV), 
pseudoscalar Higgs boson $a$ ($m_a\ge m_t$), and $Z^\prime$ 
($m_{Z^\prime} \ge m_t$) contribute to the top-quark pair production, 
we could not find  the favored parameter regions which can be accommodated 
with the present  experiments in the 1$\sigma$ level.
In this case, we must relax the experimental bound to the 2$\sigma$ level,
or consider the light $Z^\prime$ model where the lightest Higgs boson
($\sim 125$ GeV) has a negligible off-diagonal Yukawa coupling and
the second lightest Higgs boson has a large off-diagonal Yukawa coupling.

Usually the original $Z^\prime$ model is considered having been disfavored, 
independently by the stringent upper bound on the same-sign top-quark pair 
production cross sections and the shape of $m_{t\bar{t}}$ spectrum 
at high $m_{t\bar{t}}$ region, 
as well as the null results of the charge asymmetry at the LHC.   
However, our current study and the previous analyses~\cite{u1models,u1models2} 
show that these conclusions are too premature, since 
extra Higgs doublets charged under U(1)$^\prime$  are required for realistic 
Yukawa couplings and they usually contribute to the top physics.
When we extend the Higgs sector and generate the top quark mass correctly
as presented in Refs.~\cite{u1models,u1models2},  there appear 
the neutral Higgs mediated FCNC in the up-quark sector, which contributes
to all the observables related with the top FB asymmetry at the Tevatron and 
the charge asymmetry at the LHC. 
It is neither consistent nor complete to do phenomenological analysis, 
keeping only the $Z^\prime$ contribution with the physical top mass equal 
to the experimental value. Our predictions on the correlations among 
$A_{\rm FB}^t$,  $\sigma_{tt}$ and $A_{\rm C}^y$ are completely different 
from the results in the literature (see Ref.~\cite{AguilarSaavedra:2011ug}, 
for example).  These comments about new Higgs doublets would apply 
to other models where new spin-1 vector bosons have chiral couplings 
to the SM fermions. 

Also it should be emphasized that there is no mathematical limit 
where one can integrate out the U(1)$^\prime$-charged new Higgs doublets 
assuming they are very heavy, and keep only the light $Z^\prime$. 
The reason is that the would-be Goldstone bosons eaten 
by the SM $W_L$ and $Z_L$ are shared among all the Higgs doublets with 
nonzero VEV's. Since these massless components are also reside 
in the U(1)$^\prime$-charged Higgs doublets, one can not integrate them out.  
Also there would be the problem with the realistic Yukawa couplings 
for the up-type quarks if one would have integrated them out.
On the other hand, one can consider the opposite limit where 
the $Z^\prime$ boson is very heavy and can be integrated out. 
Then the low energy effective theory would be 
multi-Higgs doublet models with some FCNC interactions to the up-type quarks.
Eventually our model may be excluded by the future experiments 
on the observables we studied in this paper or another observables. 
However the $Z^\prime$ model with U(1)$^\prime$-flavored Higgs doublets 
are still viable explanations to the top FB asymmetry at the Tevatron, 
without conflict with the same-sign top-quark pair productions, 
the $m_{t\bar{t}}$ 
distributions, or the charge asymmetry at the LHC, unlike the common belief.

Finally, it should be mentioned that our study is not complete yet, 
because there are extra fields which we simply assumed to be subdominant for 
physics of $A_\textrm{FB}^t$.
In this paper, we included the heavier neutral  scalar bosons and could achieve 
the large $A_\textrm{FB}^t$ without too large same-sign top-quark pair cross section. 
We demonstrated the lighter neutral scalar could correspond to the excess at LHC. 
Such heavier neutral scalar bosons should also be constrained by the Higgs 
search at the LHC and Tevatron. 

\vspace{0.5cm}
{\it Note Added}

While we are finalizing this paper, both the ATLAS and CMS collaborations
have announced the discover of a Higgs-like scalar boson with  $\sim 125$ GeV 
mass~\cite{atlashiggs,cmshiggs}.  They have observed a rather larger excess in the 
$h \to \gamma\gamma$ channel  and smaller signals in the $h\to WW/ZZ$ channels.  
If the branching ratios settle down at the present values, our models will severely be 
constrained.  We anticipate further results of this Higgs-like scalar boson.

Furthermore, the CMS collaboration has announced more stringent bound
on the cross section for the same-sign top-quark pair production 
at the LHC: $\sigma^{tt} \le 0.39$ pb at 95\%
confidence level~\cite{Chatrchyan:2012sa,cmsnewtt}.
This strong bound would exclude the light $Z^\prime$ boson case.
However, we note that the light scalar Higgs boson model 
with $Z^\prime$, $H$, and $a$, where $m_h=125$ GeV,
$m_{Z^\prime}\ge 160$ GeV and $m_{H,a}\ge m_t$
might be in agreement with this upper bound in a certain parameter space.



\begin{acknowledgments}
We thank Korea Institute for Advanced Study for providing computing resources 
(KIAS Center for Advanced Computation Abacus System) for this work.
This work is supported in part by Basic Science Research Program
through NRF 2011-0022996 (CY), by NRF Research Grant
2012R1A2A1A01006053 (PK and CY), and by SRC program of NRF
Grant No. 20120001176 through
Korea Neutrino Research Center at Seoul National University (PK).

\end{acknowledgments}


\vspace{-1ex}

\begin{thebibliography}{999}

\bibitem{CDFlepjet}
  T.~Aaltonen {\it et al.}  [CDF Collaboration],
  Phys.\ Rev.\  D {\bf 83}, 112003 (2011)
  [arXiv:1101.0034 [hep-ex]].

\bibitem{CDFdilep}
  CDF Collaboration,
  CDF note {\bf 10436} (2011).

\bibitem{D0lepjet}
  V.~M.~Abazov {\it et al.}  [D0 Collaboration],
  Phys.\ Rev.\ D {\bf 84}, 112005 (2011)  
  [arXiv:1107.4995 [hep-ex]].  

\bibitem{cdfnew}
  CDF Collaboration,
  CDF Conf. note {\bf 10807} (2012).

\bibitem{smnlo}
  O.~Antunano, J.~H.~Kuhn and G.~Rodrigo,
  Phys.\ Rev.\  D {\bf 77}, 014003 (2008)
  [arXiv:0709.1652 [hep-ph]].

\bibitem{smnlo2}
  V.~Ahrens, A.~Ferroglia, M.~Neubert, B.~D.~Pecjak and L.~L.~Yang,
  Phys.\ Rev.\  D {\bf 84}, 074004 (2011)
  [arXiv:1106.6051 [hep-ph]].

\bibitem{smewnlo}
  W.~Hollik and D.~Pagani,
  Phys.\ Rev.\  D {\bf 84}, 093003 (2011)
  [arXiv:1107.2606 [hep-ph]].

\bibitem{smewnlo2}
  J.~H.~Kuhn and G.~Rodrigo,
  arXiv:1109.6830 [hep-ph].


\bibitem{Choudhury:2007ux}
  D.~Choudhury, R.~M.~Godbole, R.~K.~Singh and K.~Wagh,
  Phys.\ Lett.\  B {\bf 657}, 69 (2007)
  [arXiv:0705.1499 [hep-ph]].


\bibitem{Jung:2009jz}
  S.~Jung, H.~Murayama, A.~Pierce and J.~D.~Wells,
  Phys.\ Rev.\  D {\bf 81}, 015004 (2010)
  [arXiv:0907.4112 [hep-ph]].

\bibitem{Cheung:2009ch}
  K.~Cheung, W.~Y.~Keung and T.~C.~Yuan,
  Phys.\ Lett.\  B {\bf 682}, 287 (2009)
  [arXiv:0908.2589 [hep-ph]].

\bibitem{Shu:2009xf}
  J.~Shu, T.~M.~P.~Tait and K.~Wang,
  Phys.\ Rev.\  D {\bf 81}, 034012 (2010)
  [arXiv:0911.3237 [hep-ph]].

\bibitem{Arhrib:2009hu}
  A.~Arhrib, R.~Benbrik and C.~H.~Chen,
  Phys.\ Rev.\  D {\bf 82}, 034034 (2010)
  [arXiv:0911.4875 [hep-ph]].

\bibitem{Dorsner:2009mq}
  I.~Dorsner, S.~Fajfer, J.~F.~Kamenik and N.~Kosnik,
  Phys.\ Rev.\  D {\bf 81}, 055009 (2010)
  [arXiv:0912.0972 [hep-ph]].

\bibitem{Jung:2009pi}
  D.~W.~Jung, P.~Ko, J.~S.~Lee and S.~h.~Nam,
  Phys.\ Lett.\  B {\bf 691}, 238 (2010)
  [arXiv:0912.1105 [hep-ph]].

\bibitem{Barger:2010mw}
  V.~Barger, W.~Y.~Keung and C.~T.~Yu,
  Phys.\ Rev.\  D {\bf 81}, 113009 (2010)
  [arXiv:1002.1048 [hep-ph]].

\bibitem{Cao:2010zb}
  Q.~H.~Cao, D.~McKeen, J.~L.~Rosner, G.~Shaughnessy and C.~E.~M.~Wagner,
  Phys.\ Rev.\  D {\bf 81}, 114004 (2010)
  [arXiv:1003.3461 [hep-ph]].

\bibitem{Xiao:2010hm}
  B.~Xiao, Y.~k.~Wang and S.~h.~Zhu,
  Phys.\ Rev.\  D {\bf 82}, 034026 (2010)
  [arXiv:1006.2510 [hep-ph]].

\bibitem{Jung:2010yn}
  D.~W.~Jung, P.~Ko and J.~S.~Lee,
  Phys.\ Lett.\  B {\bf 701}, 248 (2011)
  [arXiv:1011.5976 [hep-ph]].

\bibitem{Choudhury:2010cd}
  D.~Choudhury, R.~M.~Godbole, S.~D.~Rindani and P.~Saha,
  Phys.\ Rev.\  D {\bf 84}, 014023 (2011)
  [arXiv:1012.4750 [hep-ph]].

\bibitem{Cheung:2011qa}
  K.~Cheung and T.~C.~Yuan,
  Phys.\ Rev.\  D {\bf 83}, 074006 (2011)
  [arXiv:1101.1445 [hep-ph]].

\bibitem{Gresham:2011dg}
  M.~I.~Gresham, I.~W.~Kim and K.~M.~Zurek,
  Phys.\ Rev.\  D {\bf 84}, 034025 (2011)
  [arXiv:1102.0018 [hep-ph]].

\bibitem{Bhattacherjee:2011nr}
  B.~Bhattacherjee, S.~S.~Biswal and D.~Ghosh,
  Phys.\ Rev.\  D {\bf 83}, 091501 (2011)
  [arXiv:1102.0545 [hep-ph]].

\bibitem{Barger:2011ih}
  V.~Barger, W.~Y.~Keung and C.~T.~Yu,
  Phys.\ Lett.\  B {\bf 698}, 243 (2011)
  [arXiv:1102.0279 [hep-ph]].

\bibitem{Grinstein:2011yv}
  B.~Grinstein, A.~L.~Kagan, M.~Trott and J.~Zupan,
  Phys.\ Rev.\ Lett.\  {\bf 107}, 012002 (2011)
  [arXiv:1102.3374 [hep-ph]].

\bibitem{Patel:2011eh} 
  K.~M.~Patel and P.~Sharma,
  JHEP {\bf 1104}, 085 (2011)  
  [arXiv:1102.4736 [hep-ph]].  

\bibitem{Isidori:2011dp}
  G.~Isidori and J.~F.~Kamenik,
  Phys.\ Lett.\  B {\bf 700}, 145 (2011)
  [arXiv:1103.0016 [hep-ph]].

\bibitem{Zerwekh:2011wf}
  A.~R.~Zerwekh,
  Phys.\ Lett.\  B {\bf 704}, 62 (2011)
  [arXiv:1103.0956 [hep-ph]].

\bibitem{Barreto:2011au}
  E.~R.~Barreto, Y.~A.~Coutinho and J.~Sa Borges,
  Phys.\ Rev.\  D {\bf 83}, 054006 (2011)
  [arXiv:1103.1266 [hep-ph]].

\bibitem{Foot:2011xu}
  R.~Foot,
  Phys.\ Rev.\  D {\bf 83}, 114013 (2011)
  [arXiv:1103.1940 [hep-ph]].

\bibitem{Ligeti:2011vt}
  Z.~Ligeti, G.~M.~Tavares and M.~Schmaltz,
  JHEP {\bf 1106}, 109 (2011)
  [arXiv:1103.2757 [hep-ph]].

\bibitem{AguilarSaavedra:2011vw}
  J.~A.~Aguilar-Saavedra and M.~Perez-Victoria,
  JHEP {\bf 1105}, 034 (2011)
  [arXiv:1103.2765 [hep-ph]].

\bibitem{Gresham:2011pa}
  M.~I.~Gresham, I.~W.~Kim and K.~M.~Zurek,
  Phys.\ Rev.\  D {\bf 83}, 114027 (2011)
  [arXiv:1103.3501 [hep-ph]].

\bibitem{Shu:2011au}
  J.~Shu, K.~Wang and G.~Zhu,
  Phys.\ Rev.\  D {\bf 85}, 034008 (2012)
  [arXiv:1104.0083 [hep-ph]].

\bibitem{AguilarSaavedra:2011zy}
  J.~A.~Aguilar-Saavedra and M.~Perez-Victoria,
  Phys.\ Lett.\  B {\bf 701}, 93 (2011)
  [arXiv:1104.1385 [hep-ph]].

\bibitem{Nelson:2011us}
  A.~E.~Nelson, T.~Okui and T.~S.~Roy,
  Phys.\ Rev.\  D {\bf 84}, 094007 (2011)
  [arXiv:1104.2030 [hep-ph]].

\bibitem{Jung:2011ua}
  S.~Jung, A.~Pierce and J.~D.~Wells,
  Phys.\ Rev.\  D {\bf 84}, 055018 (2011)
  [arXiv:1104.3139 [hep-ph]].

\bibitem{Zhu:2011ww}
  G.~Zhu,
  Phys.\ Lett.\  B {\bf 703}, 142 (2011)
  [arXiv:1104.3227 [hep-ph]].

\bibitem{Jung:2011ue}
  D.~W.~Jung, P.~Ko and J.~S.~Lee,
  Phys.\ Rev.\  D {\bf 84}, 055027 (2011)
  [arXiv:1104.4443 [hep-ph]].

\bibitem{Babu:2011yw}
  K.~S.~Babu, M.~Frank and S.~K.~Rai,
  Phys.\ Rev.\ Lett.\  {\bf 107}, 061802 (2011)
  [arXiv:1104.4782 [hep-ph]].

\bibitem{Krohn:2011tw}
  D.~Krohn, T.~Liu, J.~Shelton and L.~T.~Wang,
  Phys.\ Rev.\  D {\bf 84}, 074034 (2011)
  [arXiv:1105.3743 [hep-ph]].

\bibitem{Hektor:2011ms}
  A.~Hektor, G.~Hutsi, M.~Kadastik, K.~Kannike, M.~Raidal and D.~M.~Straub,
  Phys.\ Rev.\  D {\bf 84}, 031701 (2011)
  [arXiv:1105.5644 [hep-ph]].

\bibitem{Cui:2011xy}
  Y.~Cui, Z.~Han and M.~D.~Schwartz,
  JHEP {\bf 1107}, 127 (2011)
  [arXiv:1106.3086 [hep-ph]].

\bibitem{Gabrielli:2011jf}
  E.~Gabrielli and M.~Raidal,
  Phys.\ Rev.\  D {\bf 84}, 054017 (2011)
  [arXiv:1106.4553 [hep-ph]].

\bibitem{Duraisamy:2011pt}
  M.~Duraisamy, A.~Rashed and A.~Datta,
  Phys.\ Rev.\  D {\bf 84}, 054018 (2011)
  [arXiv:1106.5982 [hep-ph]].

\bibitem{AguilarSaavedra:2011ug}
  J.~A.~Aguilar-Saavedra and M.~Perez-Victoria,
  JHEP {\bf 1109}, 097 (2011)
  [arXiv:1107.0841 [hep-ph]].

\bibitem{Tavares:2011zg}
  G.~M.~Tavares and M.~Schmaltz,
  Phys.\ Rev.\  D {\bf 84}, 054008 (2011)
  [arXiv:1107.0978 [hep-ph]].

\bibitem{Vecchi:2011ab}
  L.~Vecchi,
  JHEP {\bf 1110}, 003 (2011)
  [arXiv:1107.2933 [hep-ph]].

\bibitem{Shao:2011wa}
  D.~Y.~Shao, C.~S.~Li, J.~Wang, J.~Gao, H.~Zhang and H.~X.~Zhu,
  Phys.\ Rev.\  D {\bf 84}, 054016 (2011)
  [arXiv:1107.4012 [hep-ph]].

\bibitem{Blum:2011fa}
  K.~Blum, Y.~Hochberg and Y.~Nir,
  JHEP {\bf 1110}, 124 (2011)
  [arXiv:1107.4350 [hep-ph]].

\bibitem{Gresham:2011fx}
  M.~I.~Gresham, I.~W.~Kim and K.~M.~Zurek,
  Phys.\ Rev.\  D {\bf 85}, 014022 (2012)
  [arXiv:1107.4364 [hep-ph]].

\bibitem{Frank:2011rb}
  M.~Frank, A.~Hayreter and I.~Turan,
  Phys.\ Rev.\  D {\bf 84}, 114007 (2011)
  [arXiv:1108.0998 [hep-ph]].

\bibitem{Davoudiasl:2011tv}
  H.~Davoudiasl, T.~McElmurry and A.~Soni,
  Phys.\ Rev.\  D {\bf 85}, 054001 (2012)
  [arXiv:1108.1173 [hep-ph]].

\bibitem{Jung:2011id}
  S.~Jung, A.~Pierce and J.~D.~Wells,
  Phys.\ Rev.\  D {\bf 84}, 091502 (2011)
  [arXiv:1108.1802 [hep-ph]].

\bibitem{Liu:2011dh}
  J.~Y.~Liu, Y.~Tang and Y.~L.~Wu,
  J.\ Phys.\ G {\bf 39}, 055003 (2012)  
  [arXiv:1108.5012 [hep-ph]]. 

\bibitem{Kolodziej:2011ir}
  K.~Kolodziej,
  Phys.\ Lett.\ B {\bf 710}, 671 (2012)  
  [arXiv:1110.2103 [hep-ph]].  


\bibitem{Ng:2011jv}
  J.~N.~Ng and P.~T.~Winslow,
  JHEP {\bf 1202}, 140 (2012)  
  [arXiv:1110.5630 [hep-ph]].  


\bibitem{Yan:2011tf}
  K.~Yan, J.~Wang, D.~Y.~Shao and C.~S.~Li,
  Phys.\ Rev.\  D {\bf 85}, 034020 (2012)
  [arXiv:1110.6684 [hep-ph]].

\bibitem{Jung:2011ym}
  D.~W.~Jung, P.~Ko and J.~S.~Lee,
  Phys.\ Lett.\  B {\bf 708}, 157 (2012)
  [arXiv:1111.3180 [hep-ph]].

\bibitem{Wang:2011mra}
  L.~Wang, L.~Wu and J.~M.~Yang,
  Phys.\ Rev.\ D {\bf 85}, 075017 (2012)  
  [arXiv:1111.4771 [hep-ph]].

\bibitem{Biswal:2012mr}
  S.~S.~Biswal, S.~Mitra, R.~Santos, P.~Sharma, R.~K.~Singh and M.~Won,
  Phys.\ Rev.\ D {\bf 86}, 014016 (2012)  
  [arXiv:1201.3668 [hep-ph]].

\bibitem{Ko:2012sm}
  P.~Ko,
  arXiv:1202.0367 [hep-ph].

\bibitem{Gresham:2012wc} 
  M.~I.~Gresham, I.~-W.~Kim, S.~Tulin and K.~M.~Zurek,
  Phys.\ Rev.\ D {\bf 86}, 034029 (2012)  
  [arXiv:1203.1320 [hep-ph]].

\bibitem{Grinstein:2012pn} 
  B.~Grinstein, C.~W.~Murphy, D.~Pirtskhalava and P.~Uttayarat,
  JHEP {\bf 1208}, 073 (2012)  
  [arXiv:1203.2183 [hep-ph]].  

\bibitem{Han:2012qu}
  C.~Han, N.~Liu, L.~Wu and J.~M.~Yang,
  Phys.\ Lett.\ B {\bf 714}, 295 (2012)  
  [arXiv:1203.2321 [hep-ph]].

\bibitem{Duffty:2012zz}
  D.~Duffty, Z.~Sullivan and H.~Zhang,
  JHEP {\bf 1206}, 164 (2012)  
  [arXiv:1203.4889 [hep-th]].



\bibitem{u1models}
  P.~Ko, Y.~Omura and C.~Yu,
  Phys.\ Rev.\ D {\bf 85}, 115010 (2012)  
  [arXiv:1108.0350 [hep-ph]].

\bibitem{u1models2}
  P.~Ko, Y.~Omura and C.~Yu,
  JHEP {\bf 1201}, 147 (2012)
  [arXiv:1108.4005 [hep-ph]].

\bibitem{u1models3}
  P.~Ko, Y.~Omura and C.~Yu,
  Nuovo Cim.\ C {\bf 035N3}, 245 (2012)  
  [arXiv:1201.1352 [hep-ph]].

\bibitem{Babu:2011sd}
  K.~S.~Babu, J.~Julio and Y.~Zhang,
  Nucl.\ Phys.\  B {\bf 858}, 468 (2012)
  [arXiv:1111.5021 [hep-ph]].


\bibitem{Chatrchyan:2011dk}
  S.~Chatrchyan {\it et al.}  [CMS Collaboration],
  JHEP {\bf 1108}, 005 (2011)
  [arXiv:1106.2142 [hep-ex]].

\bibitem{Aad:2012bb}
  G.~Aad {\it et al.}  [ATLAS Collaboration],
  JHEP {\bf 1204}, 069 (2012)  
  [arXiv:1202.5520 [hep-ex]].

\bibitem{Cao:2011ew}
  J.~Cao, L.~Wang, L.~Wu and J.~M.~Yang,
  Phys.\ Rev.\  D {\bf 84}, 074001 (2011)
  [arXiv:1101.4456 [hep-ph]].

\bibitem{Berger:2011ua}
  E.~L.~Berger, Q.~H.~Cao, C.~R.~Chen, C.~S.~Li and H.~Zhang,
  Phys.\ Rev.\ Lett.\  {\bf 106}, 201801 (2011)
  [arXiv:1101.5625 [hep-ph]].


\bibitem{atlasacy}
  ATLAS Collaboration,
  ATLAS-CONF-2011-106 (2011).

\bibitem{cmsacy}
  CMS Collaboration, CMS-PAS-TOP-11-030 (2011).

\bibitem{cmsmtt}
  CMS Collaboration, CMS-PAS-TOP-11-009 (2011).

\bibitem{atlasmtt}
  ATLAS Collaboration, ATLAS-CONF-2012-029 (2012).

\bibitem{cdfttbar}
  CDF Collaboration, CDF note {\bf 9913} (2009).

\bibitem{d0ttbar}
  V.~M.~Abazov {\it et al.}  [D0 Collaboration],
  Phys.\ Lett.\  B {\bf 704}, 403 (2011)
  [arXiv:1105.5384 [hep-ex]].

\bibitem{cmsttbar}
  CMS Collaboration, CMS-PAS-TOP-11-024 (2011).

\bibitem{atlasttbar}
  ATLAS Collaboration, ATLAS-CONF-2012-024 (2012).

\bibitem{cteq}
  J.~Pumplin, D.~R.~Stump, J.~Huston, H.~L.~Lai, P.~M.~Nadolsky and W.~K.~Tung,
  JHEP {\bf 0207}, 012 (2002)
  [arXiv:hep-ph/0201195].

\bibitem{ATLAS:2012ae}
  G.~Aad {\it et al.}  [ATLAS Collaboration],
  Phys.\ Lett.\  B {\bf 710}, 49 (2012)
  [arXiv:1202.1408 [hep-ex]].

\bibitem{Chatrchyan:2012tx}
  S.~Chatrchyan {\it et al.}  [CMS Collaboration],
  Phys.\ Lett.\  B {\bf 710}, 26 (2012)
  [arXiv:1202.1488 [hep-ex]].

\bibitem{atlashiggs}
  G.~Aad {\it et al.}  [ATLAS Collaboration],
  Phys.\ Lett.\ B {\bf 716}, 1 (2012)  
  [arXiv:1207.7214 [hep-ex]].

\bibitem{cmshiggs}
  S.~Chatrchyan {\it et al.}  [CMS Collaboration],
  Phys.\ Lett.\ B {\bf 716}, 30 (2012)  
  [arXiv:1207.7235 [hep-ex]].

\bibitem{Chatrchyan:2012sa} 
  S.~Chatrchyan {\it et al.}  [CMS Collaboration],
  JHEP {\bf 1208}, 110 (2012)  
  [arXiv:1205.3933 [hep-ex]].

\bibitem{cmsnewtt}
  CMS Collaboration, CMS-PAS-SUS-12-017 (2012).




\end{thebibliography}
\end{document}